\documentclass{article}
\usepackage[utf8]{inputenc}
\usepackage[left=1in,right=1in,top=1in,bottom=1in,includefoot,includehead,headheight=13.6pt]{geometry}

\usepackage{amsmath,amssymb,amsfonts}
\usepackage{algorithmic}
\usepackage{textcomp}
\usepackage{graphicx}
\usepackage{booktabs}
\usepackage{multirow}
\usepackage{xcolor}
\usepackage{subfig}

\usepackage{xcolor}
\definecolor{darkolivegreen}{rgb}{0.33, 0.42, 0.18}
\definecolor{celestialblue}{rgb}{0.29, 0.59, 0.82}

\usepackage[colorlinks,
            linkcolor=darkolivegreen,
            citecolor=darkolivegreen,
            urlcolor=darkolivegreen]{hyperref}

\hypersetup{pdftitle={CGTJS}, pdfauthor={Felipe Galarce Marin}, 
}

\usepackage{fullpage}
\usepackage{amsmath, amsthm, amssymb}
\usepackage{thmtools, thm-restate}
\usepackage{dsfont}
\usepackage{mathtools}
\mathtoolsset{showonlyrefs=true}
\usepackage{graphicx}   
\usepackage{color}
\usepackage{setspace}
\theoremstyle{definition}

\numberwithin{equation}{section} 

\newcommand{\bR}{\ensuremath{\mathbb{R}}}

\def\[{\left[}
\def\]{\right]}
\def\<{\langle}
\def\>{\rangle}
\def\({\left(}
\def\){\right)}
\def\[{\left [}
\def\]{\right]}
\def\({\left(}
\def\){\right)}

\usepackage{enumitem}

\newcommand{\bu}{\boldsymbol{u}}
\newcommand{\bn}{\boldsymbol{n}}
\newcommand{\inlet}{\Gamma_{\text{in}}}
\newcommand{\outlet}{\Gamma_{\text{out}}}
\newcommand{\wall}{\Gamma_{\text{wall}}}
\newcommand{\review}[1]{#1}

\newcommand{\wss}{
     \boldsymbol{l} \times \left\{ \left(2\mu \dot{\boldsymbol{\varepsilon}}\boldsymbol{l} \right)\times \boldsymbol{l} \right\}}

\def\[{\left[}
\def\]{\right]}
\def\<{\langle}
\def\>{\rangle}
\def\({\left(}
\def\){\right)}
\def\[{\left [}
\def\]{\right]}
\def\({\left(}
\def\){\right)}

\begin{document}

\title{Evaluating the Impact of Blood Rheology in Hemodynamic Parameters by 4D Flow MRI in Large Vessels Considering the Hematocrit Effect}

\author{Hernan Mella\footnote{School of Electrical Engineering, Pontificia Universidad Católica de Valparaíso, Chile.}, Felipe Galarce$^0$\footnote{School of Civil Engineering, Pontificia Universidad Católica de Valparaíso, Chile.}, Tetsuro Sekine\footnote{Department of Radiology, Nippon Medical School Musashi Kosugi Hospital, Japan.}, Julio Sotelo\footnote{Departamento de Informática, Universidad Técnica Federico Santa María, Chile}, Ernesto Castillo \footnote{Department of Mechanical Engineering, Universidad de Santiago de Chile. Chile}
}
\maketitle

\begin{center}
\textbf{Abstract}\\
\end{center}
Aortic hemodynamic parameters estimated from 4D Flow Magnetic Resonance (MR) velocity measurements are often estimated using a constant Newtonian viscosity, neglecting blood's shear-thinning behavior. \review{The aim of this work is to estimate and assess whether Newtonian viscosity is sufficient to quantify these parameters, given the non-Newtonian nature of blood. Additionally, we demonstrate that shear-thinning effects remain observable in large vessels despite artifacts commonly present in 4D Flow MR images.} To address this, we quantified the impact of blood rheology and hematocrit (Hct) on Wall Shear Stress (WSS), the rate of viscous Energy Loss ($\dot{E}_L$), and the Oscillatory Shear Index (OSI) based on velocity data obtained from 4D Flow MR images. Using a Hct-dependent power-law non-Newtonian model with experimentally derived rheological parameters, we analyzed these metrics across a broad range of Hct values at physiological temperatures in both in-silico and in-vivo MR datasets. 

The results reveal significant differences between Newtonian and non-Newtonian models. In in-silico experiments, WSS and $\dot{E}_L$ differed by up to +189\% and +112\% at systole, with reductions of -74\% and -80\% at diastole, respectively, while OSI differences ranged from -23\% to -30\%. For in-vivo data, WSS and $\dot{E}_L$ deviations reached -44\% and -60\% at systole, ranging from -69\% to +73\% at diastole, with OSI differences averaging -21\%. These findings highlights the importance of accounting for non-Newtonian blood rheology when estimating hemodynamic parameters from 4D Flow MR images in large vessels, enhancing the accuracy of cardiovascular disease assessments using in-vivo aortic data.

\begin{flushright}
\small \textsl{{First two authors contributed equally}}\\
\end{flushright}


\normalsize

\section{Introduction}
\label{sec:introduction}
\review{4D (3D + time) Flow imaging is an advanced Magnetic Resonance (MR) technique that allows the measurement of spatially and temporally resolved velocity fields for any tissue \cite{doyle2022}. During the past decades, 4D flow has been extensively used to assess a wide variety of diseases by evaluating blood flow in various organs, such as the heart \cite{muraNoninvasiveLocalPulse2022,callaghan2021}, brain \cite{holmgren2020,dai2024}, and large vessels \cite{sotelo_fully_2022,sotelo3DAxialCircumferential2018}.} 

Hemodynamic parameters estimated from 4D Flow MR \review{velocity measurements} such as Wall Shear Stress (WSS), Oscillatory Shear Index (OSI), and viscous Energy Loss ($E_L$) have been successfully used for the evaluation of various cardiovascular diseases in large vessels, including Bicuspid Aortic Valve \cite{sotelo_fully_2022}, aortic coarctations \cite{jr_ladisa_computational_2011}, stenosis \cite{farag_aortic_2018}, among others \cite{Takahashi2021}. \review{The estimation of these parameters requires the use of complex image-processing methodologies and the selection of an appropriate blood viscosity definition $\mu$}, which is typically assumed to be constant,  ranging between $0.03$ to $0.045$ poise \cite{harloff_vivo_2010,bieging_vivo_2011,van_ooij_wall_2013,Barker2014,garay_assessment_2022,vantuijlHemodynamicParametersParent2024}. 

This assumption has been \review{widely} accepted in the cardiovascular MRI community for decades, serving as the gold standard for estimating viscosity-dependent hemodynamic parameters from \review{velocity measurements obtained using 4D Flow MR images}. However, blood viscosity has an intrinsic shear-thinning nonlinear behavior (i.e., viscosity decreases as shear rates increases), which, to the author's knowledge, has not been systematically considered \review{when estimating these parameters directly from blood velocity measurements obtained using 4D Flow MR imaging}.

Blood is a non-homogeneous and nonlinear \review{shear-thinning} fluid composed of red blood cells, white blood cells, and platelets suspended in plasma. Its behavior can be modeled using continuum mechanics and experimental data, resulting in non-Newtonian constitutive models. This shear-thinning behavior, which depends on the shear rate, can be effectively described by models such as the power-law \cite{Mandal2005,Mendieta2020}, Carreau \cite{Box2005,Lee2007,Gharahi2016}, Cross \cite{Abugattas2020,Mendieta2020}, and Casson models \cite{Suzuki2021}. \review{The fundamental distinction between these rheological models lies in the number of constitutive parameters required to characterize the viscosity. Among these, the power-law model is the simplest one, with only two constitutive parameters required: the power-law index parameter, which defines the rheological behavior of the fluid (shear-thinning or shear-thickening), and the consistency index.
The simplicity of the power-law model has been a key factor for the blood rheological community to propose empirical rheological-dependent/Hct correlations and Hct-dependent constitutive parameters for large ranges of red blood cell concentrations, a key issue for this article.} 

\review{Several studies have investigated the influence of Hct on blood rheology and viscosity. For instance, in 1962, Wells and Merrill \cite{wells_influence_1962} examined the role of Hct in determining blood viscosity across different shear rates, demonstrating elevated viscosities at low shear rates and reduced viscosities at high shear rates. In 1981, Quemada \cite{quemada1981} developed a viscosity model showing that increased Hct enhances Red Blood Cells (RBC) aggregation and deformation with an explicit shear-thinning behavior. In 2003, Baskurt and Meiselman \cite{baskurt2003} highlighted the exponential increase in blood viscosity with increasing Hct values due to augmented RBC aggregation, also emphasizing the blood rheology behavior. More recently, in 2020, Horner \cite{horner2020} provided comprehensive experimental data revealing that Hct is strongly correlated with viscoelastic and thixotropic behavior of blood, influencing rouleaux formation at low shear rates and RBC deformation at high shear rates. Furthermore, in 2021, Beris et al. \cite{beris2021} reinforced Hct’s central role in governing non-Newtonian behaviors, such as shear-thinning and thixotropy, through an extensive literature review.}

An increase in Hct \review{(i.e., increase in blood viscosity)} can result from conditions such as dehydration and polycythemia, while a decrease \review{(i.e., decrease in blood viscosity)} can be attributed to anemia, overhydration, kidney failure, pregnancy, or chronic inflammatory diseases \cite{mendlowitz_effect_1948,wells_influence_1962,walburn_constitutive_1976, mehri_red_2018,kundrapu_chapter_2018}. Although Hct can range from $10\%$ to $70\%$ in patients with these conditions, the normal range is $47\pm 7\%$ for men and $42\pm 6\%$ for women \cite{billett_hemoglobin_1990}.

It is well known that variations in viscosity could significantly influence the interpretation of 4D Flow MRI examinations, as changes in velocity measurements might reflect altered flow dynamics driven by viscosity variations related to underlying health conditions. Moreover, the shear rate, which acts as an independent variable in all the previously discussed rheological models, is affected by the spatial and temporal characteristics of blood flow. \review{These issues raise the question of whether the Newtonian viscosity model, which is widely accepted in the MRI community and used to estimate hemodynamic parameters directly from 4D flow measurements, is adequate for accurately characterizing the entire cardiac cycle}. In this regard, the incorporation of a non-Newtonian rheological model may lead to notable differences in hemodynamic parameters \review{estimated from 4D Flow MR images} between patients, \review{particularly under non-physiological blood composition and flow conditions. An additional consideration is whether these differences, if present, can be observed from velocity measurements obtained using 4D Flow MRI, an imaging technique that is inherently susceptible to various artifacts that can bias the estimation of these parameters \cite{bissell4DFlowCardiovascular2023}.}

\review{Whether to use Newtonian or non-Newtonian rheological models in hemodynamics remains an open question that has been extensively explored from a numerical perspective. For instance, in \cite{apostolidis2016a}, Apostolidis et al. investigated the effects of non-Newtonian rheology, represented by the Casson viscoplastic model, compared to its Newtonian counterpart, on a simplified left coronary artery geometry using Computational Fluid Dynamics (CFD) simulations. Both viscosity models were calibrated to an Hct of 40\% (representative of a healthy subject), and their findings revealed significant differences (up to 50\%) in simulation outputs, particularly in terms of WSS and peak pressure. 
Similarly, in \cite{elhanafy2019}, Elhanafy et al. studied the effects of blood shear-thinning and viscoelastic properties in an idealized abdominal aortic aneurysm using CFD. The non-Newtonian viscosity was modeled using the Carreau-Yasuda formulation, combined with the Oldroyd-B model to account for the viscoelastic component of the stress tensor, with parameters calibrated for a healthy subject. Their primary conclusion was that under low shear rate conditions, shear-thinning and viscoelastic properties of blood cannot be neglected due to the significant differences observed compared to the Newtonian model.

In \cite{fuchs2023}, Fuchs et al. evaluated the significance of non-Newtonian viscosity models, including Casson, Quemada, and Walburn-Schneck, for a fixed Hct of 45\% (representative of a healthy subject) in the thoracic aorta using CFD. Their findings were compared against simulations employing a fixed Newtonian viscosity, with differences assessed in parameters such as WSS, OSI, and time-averaged WSS (TAWSS). Similar to the findings of Apostolidis et al., their results demonstrated substantial regional variations in WSS between simulations using non-Newtonian and Newtonian viscosity models, with the largest differences observed in regions of low shear rate. However, for time- or spatially-averaged quantities such as OSI and TAWSS, these differences, while still noticeable, were reduced due to the filtering effect inherent in the averaging process. With the same idea of highlighting the importance of using non-Newtonian constitutive models, in \cite{denisco2023a}, De Nisco et al. investigated whether shear-thinning non-Newtonian rheology is necessary for accurately modeling coronary arteries. The study compared the Carreau and Newtonian viscosity models, with parameters calibrated for an Hct of 43\% (representative of a healthy subject), using CFD simulations on 144 patient-specific coronary artery geometries. The comparison focused on near-wall parameters, such as WSS, OSI, and TAWSS, as well as intravascular hemodynamic parameters, including helicity-derived metrics. Their results indicated that the assumption regarding blood rheology has a negligible impact on both WSS and helical flow profiles associated with coronary artery disease, at least for the fixed Hct considered in the study.

Regarding studies in which a wide range of hematocrits is analyzed, covering healthy and non-physiological patients,  in \cite{FARIAS2023103943}, Farias et al. conducted a numerical study using CFD simulations to investigate the influence of Hct levels (ranging from 5\% to 50\%) on blood dynamics. They employed a power-law viscosity model in a simplified carotid artery geometry, and their results were compared against a fixed Newtonian viscosity model representative of a healthy subject. The study revealed notable differences in flow dynamics across various Hct levels, evaluated in terms of vortex shedding frequency, velocity range, and flow bifurcation between the carotid outlet branches. Furthermore, the findings highlighted similarities between the non-Newtonian simulations and the Newtonian case within the healthy Hct range, whereas significant differences were observed at Hct levels outside this range.} \review{Similarly, in \cite{lynchEffectsNonNewtonianViscosity2022}, Lynch et al. investigated the effects of non-Newtonian blood viscosity on hemodynamic and transport metrics in patient-specific arterial and venous geometries. Their findings revealed that non-Newtonian blood behavior significantly alters flow dynamics, particularly in regions of low shear rate and recirculation, such as aneurysms and large veins. Compared to Newtonian models, non-Newtonian models exhibited lower in-plane velocity and vorticity but higher WSS. Transport analyses also demonstrated differences in scalar concentration and particle residence time, highlighting the importance of accurate rheological modeling to better understand cardiovascular mass transport and disease processes, including thrombosis and atherosclerosis.}


\review{Previous studies have also explored modeling and simulating blood flow using more complex rheological models to account for local phenomena, such as the Fahraeus effect \cite{barbee1971}. These approaches include considering a two-layered fluid composed of a biphasic layer (plasma and cells) and a plasmatic layer (only plasma) \cite{ponalagusamy2017}, or employing alternative blood models such as the Herschel–Bulkley fluid \cite{ponalagusamy2012}.}

\review{Further investigations have also focused on evaluating the impact of blood rheology in Fluid-Structure Interaction (FSI) simulations of the great vessels. For instance, Qiao et al. \cite{qiao2024} investigated the performance of the Quemada, Casson, Carreau, and Carreau-Yasuda models in simulating blood flow in healthy aortas using FSI methods. Their findings showed that instantaneous low shear strain rates (SSR $<$ 100 s$^{-1}$), where the non-Newtonian blood assumption is valid, constitute a significant proportion of the cardiac cycle and aortic wall, potentially influencing blood viscosity considerably. Furthermore, they demonstrated that high-viscosity regions, consistently observed in the descending aorta, emphasize the necessity of non-Newtonian models for accurate aortic hemodynamics simulations.}

\review{
There have also been interest in incorporating non-Newtonian viscosity models to estimate hemodynamic parameters from velocity measurements obtained using 4D Flow MRI. For instance, in \cite{vaclavuIntracranial4DFlow2018}, Vaclavu et al. evaluated the hemodynamic within the circle of Willis in children and adults with Sickle Cell Disease (SCD) using 4D Flow MR imaging and a non-Newtonian rheology model. Their results showed that SCD patients had lower Hct and viscosity, and higher velocity, flow and lumen area, with lower Endotelial Shear Stress (ESS) compared to healthy controls. When comparing only adults, they found that SCD patients shared similar velocity ranges compared to healthy controls, but lower ESS.

In \cite{riva4DFlowEvaluation2021}, Riva et al. investigated the effect of three viscosity models on the quantification of patient-specific viscous energy loss ($E_L$) from 4D Flow MRI within the left ventricle (LV): a fixed Newtonian viscosity of 3.7 cP, Newtonian viscosity adjusted to patient-specific Hct, and the Quemada model adjusted to patient-specific Hct. Their results indicated that the calculation of viscosity-dependent parameters, and consequently the characterization of LV blood energetics using 4D Flow MRI, can be influenced by the chosen viscosity model. This effect was particularly pronounced in regions with recirculation or low velocities. Although the study focused solely on LV diastolic dysfunction, the authors noted that their conclusions could also apply to other conditions characterized by increased recirculating blood in the LV, such as ischemic cardiomyopathy, or by non-pulsatile pulmonary blood flow, as observed in post-Fontan procedure cases. Furthermore, the authors emphasized the importance of considering Hct derangements, which are often associated with cardiovascular risk factors such as hypertension, diabetes, and obesity, among others.

In \cite{cheng4DFlowMRI2019}, Cheng and colleagues conducted in-silico studies involving geometries resembling Fontan patients, as well as in-vitro 4D Flow MRI phantom experiments, to evaluate the impact of shear-dependent fluid viscosity on in-vitro Fontan circulation. The investigation utilized one idealized and two patient-specific 3D-printed Fontan circuit models, which were connected to a pulsatile pump with parameters tuned to mimic cardiac circulation. The three phantoms were measured twice using 4D Flow MRI to obtain velocity maps: once with a Newtonian fluid and once with a non-Newtonian fluid resembling blood. The results demonstrated that in all three models, the non-Newtonian fluid with shear-dependent viscosity significantly altered flow patterns, power loss, flow distribution, and flow mixing compared to the Newtonian fluid. Moreover, the authors suggested that the observed differences could be associated with clinically impactful outcomes, emphasizing the importance of considering non-Newtonian fluid properties in hemodynamic analyses.

A similar study evaluated blood flow through a Bileaflet Mechanical Heart Valve (BMHV) using FSI simulations validated with in-vitro experimental data obtained through Particle Imaging Velocimetry \cite{yehInfluenceHematocritHemodynamics2019}. The study was conducted within a heart model designed to assess the performance and functionality of prosthetic heart valves. The authors found that variations in Hct significantly influence blood dynamics, particularly shear stress distribution and flow recirculation within the aortic sinus. They also demonstrated that increased Hct amplifies shear-thinning behavior, leading to non-linear changes in shear stress and altered valve leaflet dynamics, including asymmetric opening and closure patterns. These findings highlight the importance of incorporating patient-specific Hct values into computational models to improve predictions of BMHV performance and support personalized medical assessments.
}

Although there is disparity among the cited authors regarding the necessity of non-Newtonian blood modeling, it is important to note that most of their work has primarily focused on the healthy Hct range. Consequently, their findings cannot be generalized to cases involving abnormal blood composition. However, studies evaluating a wider range of Hct values consistently conclude that rheology plays a significant role, regardless of the specific non-Newtonian model used.

Despite the valuable insights provided by the aforementioned studies, several \review{open questions remain. For instance, the impact of neglecting shear-thinning blood behavior when estimating hemodynamic parameters from 4D Flow MRI velocity measurements for the assessment of blood hemodynamics remains unclear, particularly in the aorta, where non-Newtonian blood behavior is often overlooked by the MRI community. Furthermore, none of these studies have assessed whether the influence of shear-thinning blood behavior remains observable under common MR imaging conditions, such as partial volume effects, flow artifacts, and noise, when using 4D Flow MRI velocity measurements.}

\review{To address these open questions, in this investigation, we: (1) evaluated whether patient-specific Newtonian and non-Newtonian viscosities yield equivalent results in terms of WSS, $\dot{E}_L$ (rate of viscous energy loss dissipation), and OSI across different Hct levels when estimated from 4D Flow MR velocity measurements; (2) quantified the differences between these parameters when estimated using patient-specific non-Newtonian and state-of-the-art Newtonian viscosities commonly reported in the literature; and (3) showed that if blood exhibits shear-thinning behavior, this effect remains observable in 4D Flow MRI data, even in the presence of imaging artifacts under common imaging conditions.}

To achieve this, we conducted numerical and in-vivo experiments to estimate hemodynamic parameters from 4D Flow MR images using a power-law non-Newtonian viscosity model and Newtonian viscosities for blood rheology, comparing WSS, $\dot{E}_L$, and OSI estimations. This study integrates the 3D \review{Cauchy momentum equations (often loosely referred to as the Navier-Stokes equations when accounting for non-Newtonian fluids)} with 0D surrogate models for downstream flow and non-linear stress tensors to simulate shear-thinning blood behavior. \review{CFD simulations were used to generate synthetic 4D Flow MR images, which, together with in-vivo 4D Flow MRI data from five patients with Hypertrophic Cardiomyopathy, were analyzed for direct estimation of hemodynamic parameters using image processing methodologies. In both cases, results obtained using power-law viscosities adjusted to each patient's Hct (measured at physiological temperatures) were compared with those derived from Hct-fitted and state-of-the-art Newtonian viscosities commonly reported in the CMR literature.}

\review{We believe that this investigation could significantly impact the current estimation of hemodynamic parameters from MR images in clinical practice, as incorporating more realistic viscosity models may enhance the assessment of cardiovascular diseases. This is particularly relevant for patients with conditions affecting their Hct (such as those mentioned in earlier), which can further influence their blood dynamics. Finally, this study paves the way for integrating patient-specific, Hct-based non-Newtonian viscosities into future MRI-based diagnostic workflows.}

\section{Theory and methods}
\label{sec:methods}


\subsection{Mathematical modeling}
Let $\Omega \in \bR^3$ be the three-dimensional computational domain, with its two-dimensional boundaries decomposed into disjoint subsets such that $\partial \Omega = \inlet \cup_{k=1}^l \outlet^k \cup \wall$, where $\inlet$ represents the inlet and $\outlet$ refers to the outlets, and $\wall$ denotes the artery walls. The incompressible \review{Cauchy momentum} equations relate the velocity vector field ($\bu$) and the pressure ($p$) field as follows:

\begin{subequations}
    \begin{alignat}{2}
        \rho \frac{\partial \bu}{\partial t} + \rho \bu \cdot \nabla \bu + \nabla p - \nabla \cdot \tau &= 0 && \quad\text{  in } \Omega, \\
        \nabla \cdot \bu &= 0 &&\quad\text{ in } \Omega ,\\
        \bu &= \boldsymbol{g}(x,t) &&\quad\text{ on } \inlet, \label{eq:Ns_nn_inlet_bc}\\
        \bu &= 0 && \quad \text{ on } \wall, \\
        (\tau - p I ) \cdot \bn &= -p^k_{\text{wk}} \bn && \quad \text{ on } \outlet^k,
    \end{alignat}    
    \label{eq:NS_nn}    
\end{subequations}

\noindent where $k=1,\ldots,l$ is the outlet index, $\tau$ denotes the shear stress tensor of the fluid, and $\rho$ is the fluid density. The physiological value $\rho = 1.06$ gr$/$cm$^3$ was adopted in this study. Modeling choices for $\tau$ are discussed in the following section. Additionally, $g(x,t)$ is a is a spatially and temporally dependent boundary condition used to define the pulsatile behavior of blood flow.

\begin{figure}[!htb]
    \centering
    \includegraphics[width=0.5\textwidth]{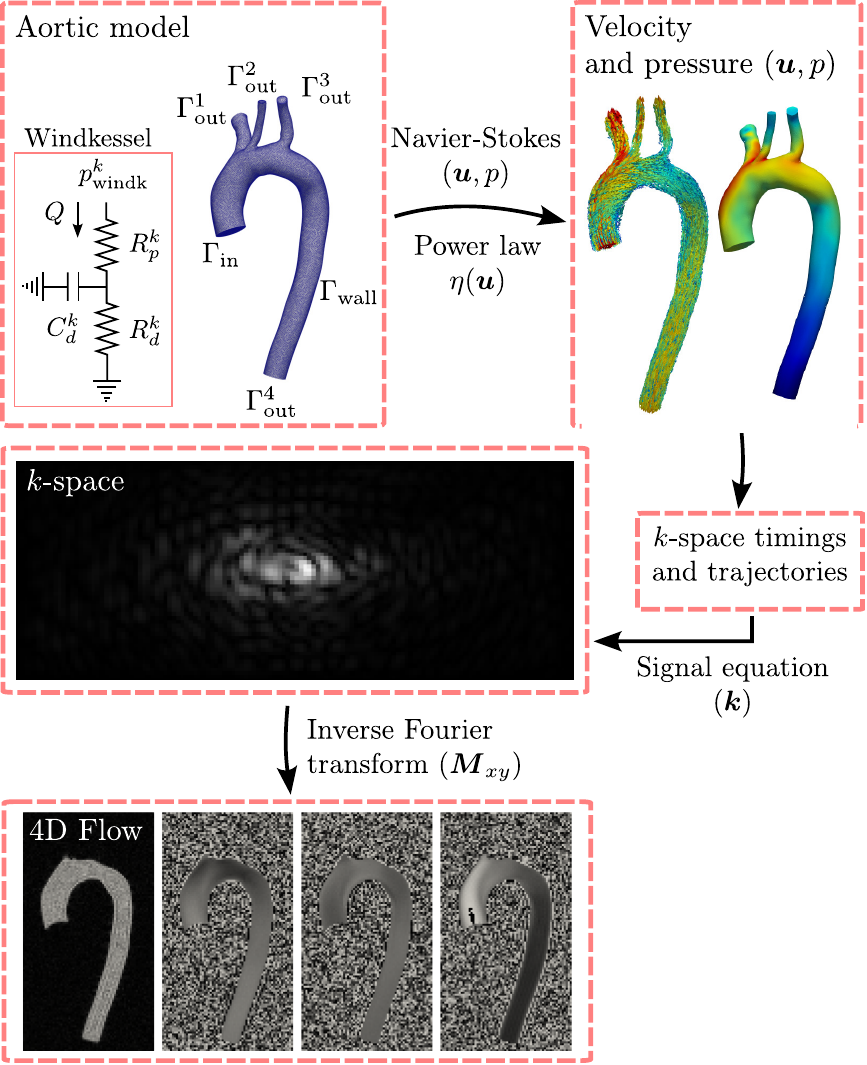}
    \caption{Simulation workflow for generating 4D Flow MR images. First, the aortic model serves as input to the CFD solver, where the \review{Cauchy momentum} equations with nonlinear viscosity are solved. Subsequently, $k$-space locations and sequence timings are generated using parameters from Table \ref{tab:mri_parameters}, and the signal equation in \eqref{eq:signal_equation} is evaluated. Trapezoidal MRI gradients are considered for the generation of sequence timings.}
    \label{fig:workflow}
\end{figure}

To model the downstream fluid, we coupled the 3D \review{Cauchy momentum} equations with reduced 0D three-parameters Windkessel models \cite{Formagia_Cardiovascular_Mathematics} (see Figure \ref{fig:workflow}), as commonly done in the literature \cite{nolte2022, GGLM2021, GLM2021, ARBIA2016175}. This approach reduces to solving an Ordinary Differential Equation (ODE) for each outlet domain:

\begin{subequations}
    \begin{alignat}{2}
        &p^k_{\text{wk}} = R_p Q^k (\bu) + p^k_d, \quad k=1,\ldots,l, \\
        &C^k \frac{d p^k_d}{d t} + \frac{p^k_d}{R^k_d} = Q^k (\bu), \quad k=1,\ldots,l.
    \end{alignat}    
\end{subequations}

\noindent where $Q^k(\bu) = \int_{\outlet^k} \bu \cdot \bn ~ds$, and $R^k_d$, $C^k_d$, and $R^k_p$, are the distal resistance, capacitance, and proximal resistance, respectively. The ODE was closed with an appropriate initial condition $p^k_{d,0}$. This configuration reproduces the downstream vascular tree and interacts with the 3D \review{Cauchy momentum} model by imposing an inwards force. It is important to note that the 0D parameters used in this investigation were calibrated within a Newtonian context.

\subsection{On the choice for the shear stress blood behavior}
\label{ssec:shear-stress-behavior}


In this work, blood rheology was modeled using both a Newtonian and a non-Newtonian purely viscous shear-thinning blood behavior. \review{A Power-law curve was chosen to characterize the blood rheology, as it is both simple (i.e., depends only on two parameters) and flexible enough to reproduce experimental data at physiological ranges. In addition, there are accepted correlations for power-law curves and the Hct level, which is directly measurable through a blood test (a necessary condition for in-vivo experiments and always available in the context of a cardiovascular MRI exam). Consequently, no additional efforts are required to define the patient's blood rheology.
In the following paragraphs, both models are summarized:} 
\begin{enumerate}
    \item \textbf{Newtonian blood behavior}: This is the gold standard in hemodynamics simulations and assumes that the shear stresses are proportional to the rate of change of the fluid's velocity vector. While this assumption is widely accepted for large blood vessels where the Reynolds number is also large, its applicability for systolic and diastolic conditions has not yet been thoroughly assessed using modern 4D Flow MRI. The model for a Newtonian fluid involves solving equations \eqref{eq:NS_nn}, where the shear stress is defined as $\tau = 2 \mu_{\text{N}} \dot{\boldsymbol{\varepsilon}}$, with $\mu_{\text{N}}$ representing the constant viscosity of blood, and $\dot{\boldsymbol{\varepsilon}}$ denoting the symmetric velocity gradient, which satisfies $2 \dot{\boldsymbol{\varepsilon}} = \nabla \bu + \nabla^T \bu$.
    \item \textbf{Power-law blood behavior}: Blood is a fluid that can be characterized by shear-thinning behavior and a \review{plastic component associated with its yield stress \cite{Suzuki2021}}. The power-law model is a two-parameter constitutive law that relates shear stress to the rate of change of the fluid's velocity vector, \review{without accounting for the plastic component of blood}, as:
     \begin{equation}
        \label{eq:power_law}
        \tau = \mu_{\text{PL}} \dot{\boldsymbol{\varepsilon}}, \quad \text{with }~ \mu_{\text{PL}} = m\dot{\gamma}^{n-1}
    \end{equation}
   where $\mu_{\text{PL}}$ denotes the apparent viscosity and $\dot{\gamma} = \sqrt{2\dot{\boldsymbol{\varepsilon}} : \dot{\boldsymbol{\varepsilon}}}$ the shear rate. The power-law index parameter $n$ defines the rheological behavior of the fluid, with $n=1$ corresponding to the Newtonian model. Shear-thinning fluids like blood are characterized by $n<1$. The parameter $m$ (with units Pa$\cdot$s$^n$), known as the consistency index.
   The values of the parameters $m$ and $n$, fitted based on those proposed in \cite{walburn_constitutive_1976,mehri_red_2018,wells_influence_1962}, were derived from experimental studies and characterized at a physiological temperature of 37 $^\circ$C (see Table \ref{tab:viscosities}). 
\end{enumerate}

\review{It is important to note that inhomogeneous blood phenomena were not considered in the selection of the rheological model. For instance, the Fahraeus effect \cite{barbee1971}, plasma skimming \cite{lee2017}, and the spatial distribution of red blood cells \cite{sherwood2014} (among others) were excluded either because they do not occur in vessels with large diameters, such as the aorta and its branches, or because the rheology model used does not account for spatially varying Hct (as this is beyond the scope of this investigation). Additionally, in MRI studies (which are describe later), it is uncommon to evaluate hemodynamic parameters in aortic branches due to resolution limitations.}


\subsection{Signal model for 4D Flow images \review{and generation of synthetic MR images}}
\label{ssec:image generation}

The generation of 4D Flow MR images used as input the blood velocity $\boldsymbol{u}=(u_x, u_y, u_z)$ obtained from CFD simulations. The image generation process involved evaluating the signal equation, which depends on the magnetization at thermal equilibrium $M_0(\boldsymbol{r})$, velocity encoding sensitivity (VENC), blood velocity $u_a$ (with $a=x,y,z$), and the relaxation time of the blood $T_2^*(\boldsymbol{r})$. The signal equation is expressed as follows:
\begin{subequations}
    \label{eq:signal_equation}
    \begin{alignat}{2}
        s^a(\boldsymbol{k}(t)) &= \int_{B} M(\boldsymbol{\boldsymbol{r}}, t) p(\boldsymbol{r}) e^{-\boldsymbol{i} 2\pi \boldsymbol{k}(t)\cdot \boldsymbol{r}(t)}~d\boldsymbol{r} ,\label{eq:signal_integral}\\
        M(\boldsymbol{\boldsymbol{r}}, t) &= M_0(\boldsymbol{r}) e^{-\boldsymbol{i} \pi/\text{VENC} u_a(\boldsymbol{r})}e^{-t/T_2^*(\boldsymbol{r})},
    \end{alignat}
\end{subequations}

In this formulation, $ p(\boldsymbol{r}) $ represents the slice profile, and $ \boldsymbol{k}(t) $ denotes the Fourier space trajectory during signal acquisition, which varies based on the chosen imaging sequence. The integration domain $ B $ in the integral refers to the object being measured, in our case, the aorta. Careful consideration of hardware limitations of the scanner are required when determining the time $ t $ at which each $ \boldsymbol{k}(t) $ location is measured. Factors such as the velocity-encoding gradients, phase-encoding gradients, and readout gradients, which depend on the maximum slew rate and amplitude of the gradient system, influence the echo time. Additionally, the readout gradients are affected by the sampling frequency of the ADC, which impacts not only the timing of each $ k $-space measurement but also the occurrence of certain artifacts.

During signal measurement, the positions of spins change due to blood dynamics. This motion was accounted for in the generation of synthetic images using a first-order approximation of the spins' positions $\boldsymbol{r}(t)$ around $t_k$:
\begin{equation}
    \boldsymbol{r}(t) = \boldsymbol{r}(t_k) + \boldsymbol{u}(t_k)(t-t_k) + \mathcal{O}(t^2),
\end{equation}
where $t_k$ represents the time where the image was taken. 

No field inhomogeneities were considered during the evaluation of the signal equation, and the integral was calculated for each $k$-space location using the Finite Elements Method (FEM). All the steps described here were implemented in the \review{Finite Element MRI (FEelMRI) library (\href{https://github.com/hmella/FEelMRI}{github.com/hmella/FEelMRI})}, which has previously been used to generate synthetic strain phantoms \cite{Mella2021a, Mella2021, HmellaPyMRStrain}, \review{and was used to generate the synthetic images in this study}. The simulation workflow described in this section is illustrated in Figure \ref{fig:workflow}.

\subsection{Estimation of power-law parameters for custom Hct values and Newtonian fitted viscosities}
\label{ssec:newtonian visco estimation}

\begin{figure}[!ht]
    \centering
    \subfloat[]{\includegraphics[width=0.5\textwidth]{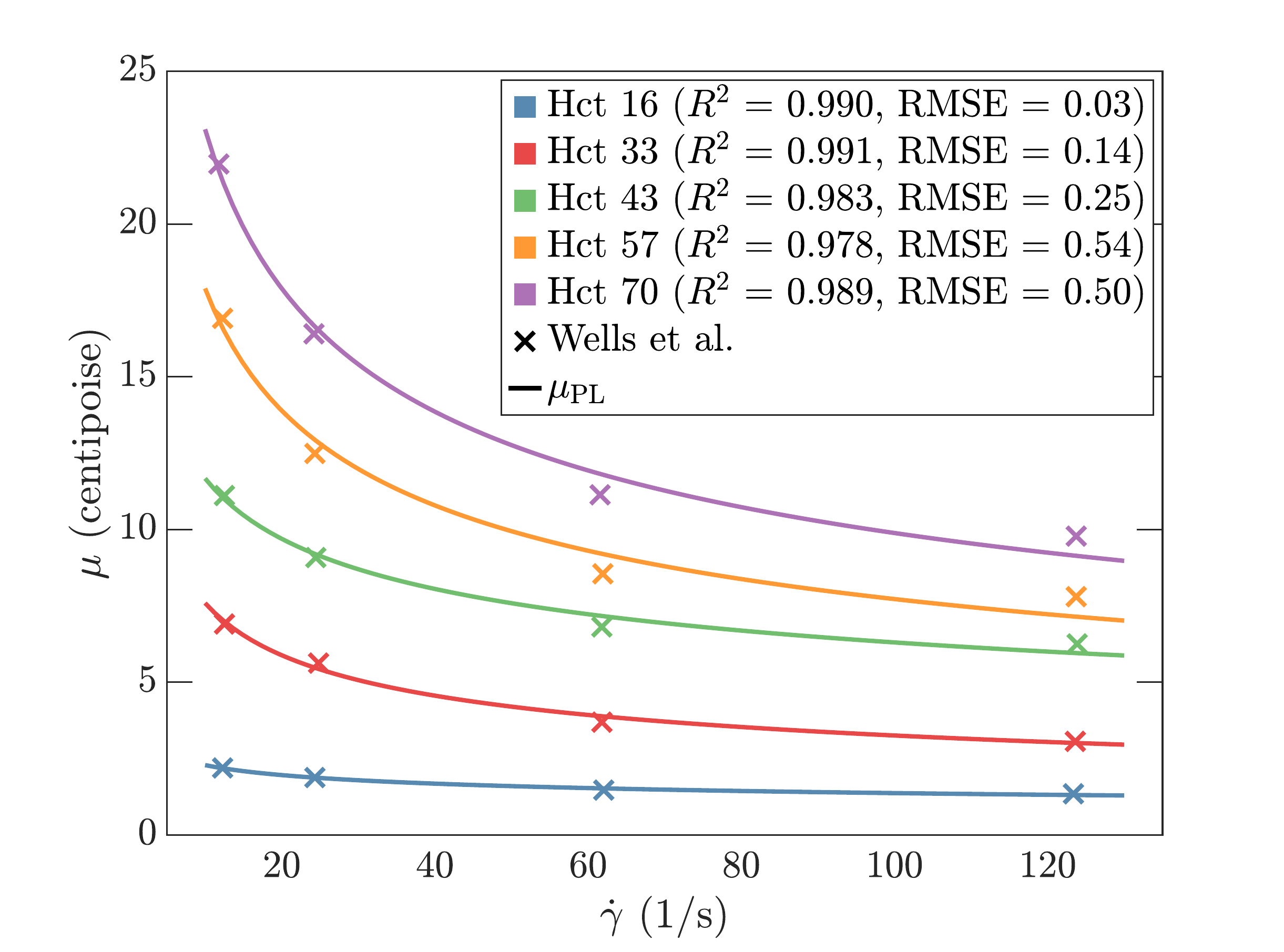}\label{fig:viscosities-a}} \hfill
    \subfloat[]{\includegraphics[width=0.5\textwidth]{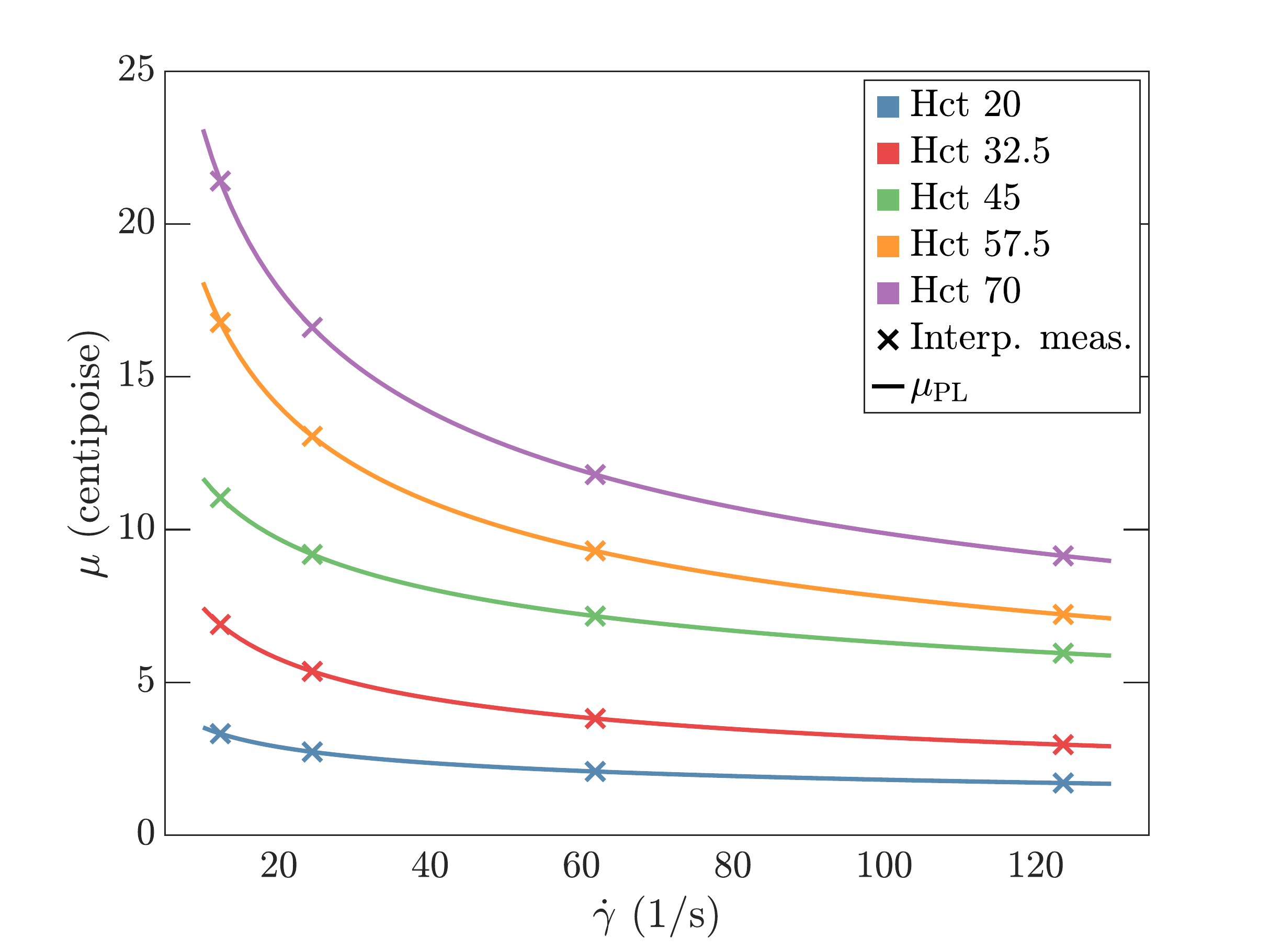}\label{fig:viscosities-b}}
    \caption{Viscosities obtained from Wells et al. \cite{wells_influence_1962}. (a) Experimental measurements and the adjusted power-law viscosities ($\mu_{\mathrm{PL}}$) derived from the data. (b) Synthetic measurements (obtained using one-dimensional linear interpolation) and power-law fits ($\mu_{\mathrm{PL}}$) for intermediate Hct levels not reported in \cite{wells_influence_1962}. Colors represents different Hct levels and, for each level, \review{the coefficient of determination ($R^2$) and root-mean-squared error (RMSE) are reported.}}
    \label{fig:viscosities}
\end{figure}

\review{Experimental} viscosity measurements at 37$^o$ for Hct values of 16\%, 33\%, 43\%, 57\%, and 70\%, across a shear rate range of 12 to 123 s$^{-1}$, were obtained from \cite{wells_influence_1962} (crosses in Figure \ref{fig:viscosities-a}). The parameters $m$ and $n$ in the power-law model described in \eqref{eq:power_law} \review{were estimated from these experimental data using a weighted least-squares fitting method to avoid data over-representation}.\review{To determine $m$ and $n$ values for any Hct within the range 16\% to 70\%, we assumed that the fitted power-law viscosity curves (see Figure \ref{fig:viscosities-a}) varied linearly between the given Hct values. 

The procedure for obtaining intermediate Hct viscosity curves was as follows:
\begin{enumerate}
    \item For a given intermediate Hct value, we identified the two fitted curves between which this value lies.
    \item Intermediate synthetic experimental measurements (crosses in Figure \ref{fig:viscosities-b}), at the same shear rates as the original experimental data, were generated using linear interpolation between the two curves.
    \item The intermediate synthetic experimental measurements were then used to fit the parameters $m$ and $n$ using a weighted least-squares fitting method (curves in Figure \ref{fig:viscosities-b}).
\end{enumerate}
It is important to note that our decision to follow this procedure was based on the fact that we could not assume $n$ varies linearly across Hct, as no clear trend was observed between these parameters (see Table \ref{tab:viscosities}). This procedure was applied to obtain the power-law parameters at Hct values of 20, 32.5, 45, 57.5, and 70\% (see Table \ref{tab:viscosities}). 

For the Newtonian viscosities fitted to Hct values, an analytical and physically consistent method designed to preserve the average shear stress within a specific shear rate range for a given Hct value was adopted. This method can be directly applied to a power-law viscosity curve to compute a corresponding Newtonian viscosity, as follows:
\begin{equation}
    \mu_{\text{NF}} = \frac{1}{\dot{\gamma}_1 - \dot{\gamma}_0}\int_{\dot{\gamma}_0}^{\dot{\gamma}_1} \mu_{\text{PL}}(\dot{\gamma})~d\dot{\gamma} = \frac{m(\dot{\gamma}_1^n - \dot{\gamma}_0^n)}{n(\dot{\gamma}_1 - \dot{\gamma}_0)}
    \label{eq:tau_mean}
\end{equation}
where $\mu_{\text{PL}}$ represents the power-law viscosity obtained using \eqref{eq:power_law}, and $\mu_{\text{NF}}$ denotes the Newtonian fitted viscosity. However, this approach yields elevated Newtonian viscosities when constrained to the limited shear rate range (12 to 123 s$^{-1}$) of the experimental measurements (see Newtonian fit 1 in Table \ref{tab:viscosities}). To mitigate this issue, a second approach extends the shear rate range of the synthetic experimental measurements to include values where blood exhibits Newtonian behavior. For instance, expanding the range to 0–2800 s$^{-1}$ results in lower Newtonian viscosities (see Table \ref{tab:viscosities}), which, for a healthy Hct of 45\%, align with values commonly reported in the literature (see Newtonian fit 2 in Table \ref{tab:viscosities}).

A third approach involves adopting the model used in \cite{riva4DFlowEvaluation2021}, which is derived by fitting a 5th-order polynomial to viscosity measurements from \cite{guyton2006textbook}. This method produces a Newtonian viscosity model that depends on Hct and can be applied to any Hct value. Additionally, it yields viscosities within the range reported in the literature (see Newtonian fit 3 in Table \ref{tab:viscosities}). However, the conditions under which these measurements were obtained remain unclear.

We chose the second approach (hereafter referred to as Newtonian fitted viscosities) for all CFD simulations of Hct-varying shear-thinning blood because, to the best of our judgment, it represents the most physically consistent translation between both models given the available data. This method ensures alignment with established values in the literature for linear models (at least for healthy Hct) while also accounting for a wide range of shear rate regimes, such as those found in the aorta \cite{sakariassen2015, panteleev2021, qiao2024}, including highly convective and even pathological scenarios.
}

Based on the above, it is important to note that three different viscosities were considered in this investigation for the generation of synthetic 4D Flow MR images:
\begin{itemize}
    \item $\mu_{\text{N}}$: The Newtonian viscosity typically used in hemodynamic contexts (not fitted by Hct values).
    \item $\mu_{\text{NF}}$: The Newtonian viscosity fitted to different Hct values.
    \item $\mu_{\text{PL}}$: The non-linear viscosity estimated using the power-law model given in \eqref{eq:power_law} and fitted to Hct values.
\end{itemize}
All of these viscosities are listed in Table \ref{tab:viscosities} for Hct values of 20, 32.5, 45, 57.5, and 70, which were used in all experiments involving CFD simulations \review{for the generation of synthetic 4D Flow MR images and the estimation of hemodynamic parameters from these data}.

\section{Experiments}
\label{sec:experiments}

\begin{table}[!t]
    \centering
    \caption{Viscosities used for CFD simulations and the estimation of hemodynamic parameters from synthetic and in-vivo 4D Flow images were obtained from \cite{wells_influence_1962} using the procedure described in Section \ref{ssec:newtonian visco estimation}. The parameters $m$ and $n$ are used to calculate $\mu_{\text{PL}}$ by evaluating Equation \eqref{eq:power_law}. \review{Emphasized column denotes the Newtonian viscosity fit used for the CFD simulations in Experiment 1.}}
    \label{tab:viscosities}
    \begin{tabular}{@{}cccccc@{}}
    \toprule
    \multirow{2}{*}{Hematocrit (\%)} & Newtonian & Power-law\footnotemark[1] & Newtonian fit 1  & \textbf{Newtonian fit 2} & Newtonian fit 3 \\
     & $\mu_\mathrm{N}$ ($10^{-3}$ Pa$\cdot$s) & $m$ ($10^{-2}$ Pa$\cdot$s$^n$), $n$ & $\mu_\mathrm{NF}$ ($10^{-3}$ Pa$\cdot$s) & $\boldsymbol{\mu_\mathrm{NF}$ ($10^{-3}}$ Pa$\cdot$s) & $\mu_\mathrm{NF}$ ($10^{-3}$ Pa$\cdot$s) \\ \midrule
    \multicolumn{6}{c}{\textit{CFD experiments}} \\ \midrule
    20.0 & \multirow{5}{*}{3.0, 3.5, 4.0, 4.5} & 0.69, 0.71 & 2.15 & \textbf{0.97} & 2.24 \\
    32.5 & {}                                  & 1.73, 0.63 & 4.00 & \textbf{1.49} & 3.26 \\
    45.0 & {}                                  & 2.42, 0.72 & 7.71 & \textbf{3.52} & 4.46 \\
    57.5 & {}                                  & 4.19, 0.64 & 9.75 & \textbf{3.64} & 6.15 \\
    70.0 & {}                                  & 5.40, 0.63 & 12.38 & \textbf{4.58} & 9.87 \\ \midrule
    \multicolumn{6}{c}{\textit{In-vivo experiments}} \\ \midrule
    28.2 & \multirow{5}{*}{3.0, 3.5, 4.0, 4.5} & 1.36, 0.65 & - & - & - \\ 
    35.2 & {}                                  & 1.83, 0.67 & - & - & - \\
    40.2 & {}                                  & 2.03, 0.71 & - & - & - \\
    46.6 & {}                                  & 2.64, 0.70 & - & - & - \\
    50.1 & {}                                  & 3.12, 0.68 & - & - & - \\ \bottomrule
    \end{tabular}
\end{table}

\subsection{Computational fluid dynamics for blood flow simulations}
\label{ssec:flow simulation}

\begin{figure}[!t]
    \centering
    \subfloat[]{\includegraphics[width=0.5\textwidth]{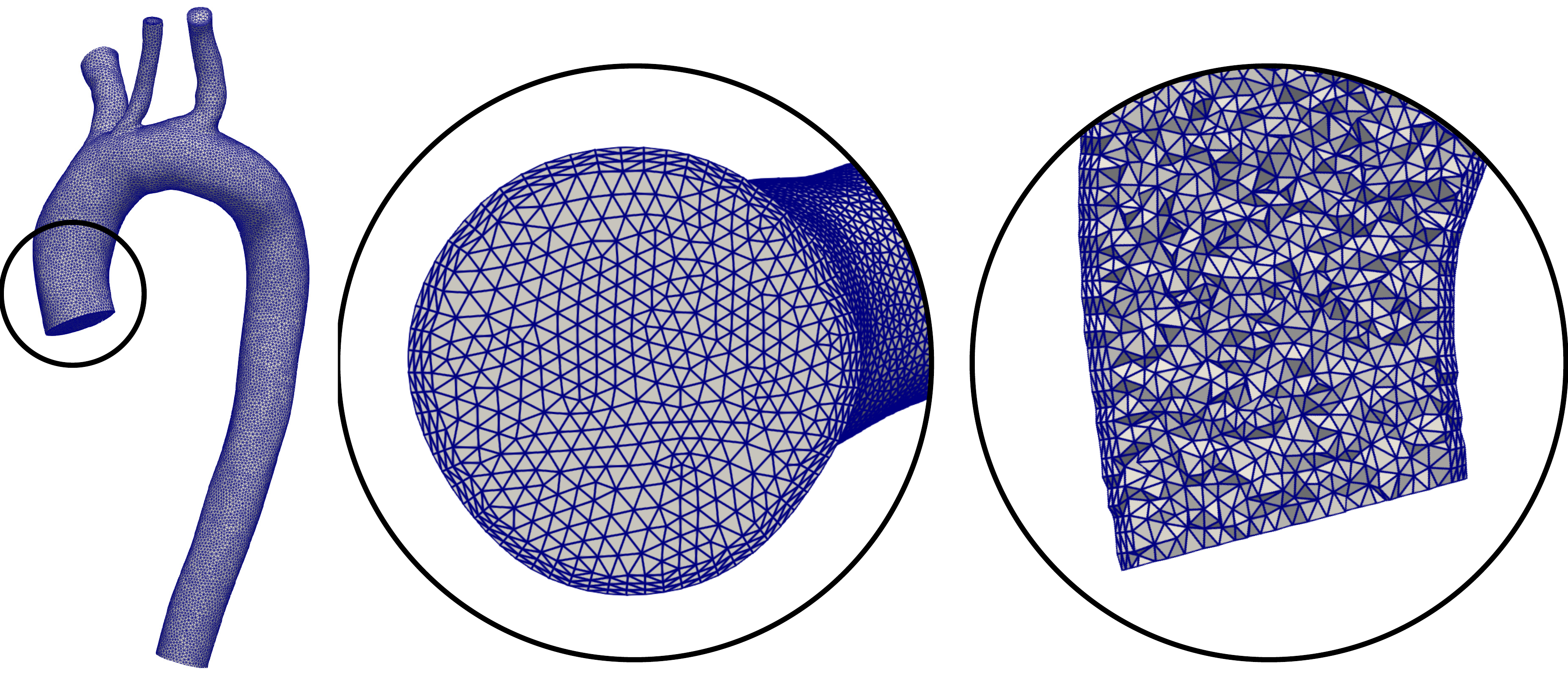}}    
    \vspace{-8pt}
    
    \subfloat[]{\includegraphics[width=0.5\textwidth]{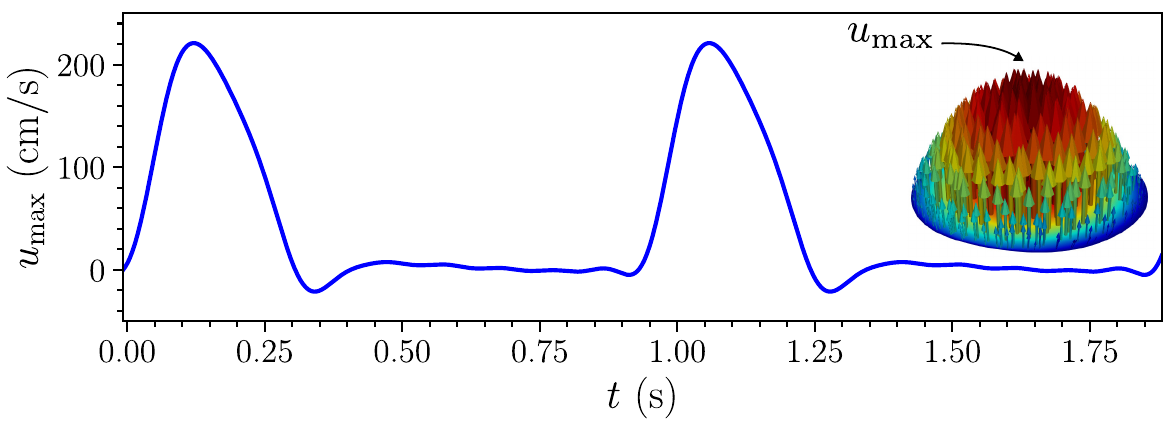}}
    \caption{(a) Near-wall mesh refinement used in CFD simulations to capture velocity gradients accurately. (b) Inlet boundary condition $u_{\text{max}}(t)$ and parabolic flow profile applied as Dirichlet boundary conditions on $\Gamma_{\text{in}}$.}
    \label{fig:inlet_bc}
    \vspace{-15pt}
\end{figure}

For all CFD experiments, the computational domain was set as a tetrahedral mesh of a patient-specific aorta model (see Figure \ref{fig:workflow}), obtained from the Vascular Model Repository \cite{jr_ladisa_computational_2011,wilson_vascular_2013,pfaller_automated_2022,vascularModel}. The mesh comprises of 127,131 vertices and 719,419 tetrahedrons and was refined near the wall to account for boundary layer effects and to accurately represent spatial gradients (see Figure \ref{fig:inlet_bc}).

The governing equations of the flow given in \eqref{eq:NS_nn} were solved using FEM with piece-wise linear Lagrange elements for $\bu$ and $p$. To overcome the well-known inf-sup constraints \cite{volkerFEM_CFD}, a Brezzi-Pitk\"aranta stabilization term was added to the formulation \cite{brezzi1984}, alongside a streamline upwind term to ensure stability for highly convective flows \cite{BROOKS1982199,TEZDUYAR19911}. Similar stabilization techniques in non-Newtonian fluids have been tested and analyzed in detail in \cite{GONZALEZ2022115586,REYES2023112086}. 
Additionally, the back-flow stabilization method proposed in \cite{esmailymoghadamComparisonOutletBoundary2011,esmailymoghadamModularNumericalMethod2013} was incorporated into all the simulations. 

The inlet boundary condition, as defined in Equation \eqref{eq:Ns_nn_inlet_bc}, is represented by $g(x,t) = u_{\text{max}}(t) \mathcal{S}(x)$. Here, $u_{\text{max}}$ was obtained from phase-contrast MRI measurements \cite{jr_ladisa_computational_2011} and the function $\mathcal{S}$ characterizes the spatial distribution of the Dirichlet condition, resembling a laminar parabolic fluid profile. Both $u_{\text{max}}$ and $\mathcal{S}$ are shown in Figure \ref{fig:inlet_bc}.

A second-order finite difference formula with a time step size of $\Delta t = 9.37 \times 10^{-4}$ s (equivalent to 1000 steps per cardiac cycle) was used for temporal discretization. Two cardiac cycles were simulated for each experiment, and only the last cycle was considered in the analysis to avoid nonphysical solutions. 

The non-linear terms in \eqref{eq:NS_nn}, i.e., the convection and the fluid constitutive, were linearized using a Picard fixed point approach, with an $H^1 \times L^2$ error tolerance of $\delta_p = 10^{-4}$. The coupling with the Windkessel model was done semi-implicitly, using a 4th-order Runge-Kutta scheme for the ODE.

The software MAD \cite{galarceThesis} was used to assemble the discrete equations and invert the linear systems. MAD is built upon the linear algebra library PETSc \cite{petsc}, and the computations are done with up to 72 CPU cores in parallel. A summary of the numerical simulation parameters is given in Table \ref{tab:fem_parameters}.

\begin{table}[!htb]
    \centering
    \caption{Simulation parameters used for CFD simulations.}
    \label{tab:fem_parameters}
    \begin{tabular}{@{}lc@{}}
        \toprule
        \multicolumn{1}{c}{FEM parameter} & Value                       \\ \midrule
        Time step ($\Delta t$) & $9.37 \times 10^{-4}$ sec      \\
       Nonlinear tolerance ($\delta_p$) & $10^{-4}$                      \\
        Backflow parameter ($\beta$) \cite{esmailymoghadamComparisonOutletBoundary2011,esmailymoghadamModularNumericalMethod2013} & 0.2                  \\ \midrule
        {Windkessel Parameter} & $\Gamma_{\text{out}}^1$, $\Gamma_{\text{out}}^2$, $\Gamma_{\text{out}}^3$, $\Gamma_{\text{out}}^4$ \\ \midrule
        Proximal resistance ($R^k_p$) & 274, 1300, 791, 141                   \\
        Distal resistance ($R^k_d$) & 5675, 19663, 10048, 2066                  \\
        Capacitance ($C^k_d\times10^{-4}$) & 5.08, 1.4416, 2.788, 13.6904              \\
        Distal pressure ($p^k_{d,0}$) & 107325, 107325, 107325, 107325                \\ \bottomrule
    \end{tabular}
\end{table}

\subsection{Generation of synthetic 4D Flow images}
\label{ssec:synthetic-4dflow}
To generate 4D Flow MR data, imaging parameters such as voxel size, number of cardiac phases, and VENC were selected according to the 2023 4D Flow MRI consensus (see Table \ref{tab:mri_parameters}) \cite{bissell4DFlowCardiovascular2023}. Hardware parameters, including maximum gradient amplitude, slew rate, and ADC frequency bandwidth, which dictated sequence timings and image quality, were taken from \cite{Bender2013} and \cite{dillingerLimitationsEchoPlanar2020}. For each Hct, a gradient-echo image was simulated with a VENC slightly higher than the maximum velocity observed at peak systole to avoid wrapping artifacts.

To emulate realistic imaging conditions, independent and uncorrelated Gaussian noise was added to both the real and imaginary components of the signal. The noise standard deviation was set at 5.2\% of the maximum amplitude of the complex signal's magnitude, resulting in a signal-to-noise ratio (SNR) of 13.5. This value is consistent with the average SNR reported for the ascending and descending aorta at 1.5T \cite{hess2015}.

Finally, the MRI signals generated from CFD simulations, using the power-law model and constant viscosities adjusted for Hct, were reconstructed into 4D Flow images by performing the inverse Fourier transform of the signal, as defined in equation \eqref{eq:signal_equation}.

\begin{table}[!htb]
    \centering
    \caption{Simulation parameters for the generation of synthetic 4D flow MR images. These parameters gave a echo-time of 1.66 ms.}
    \label{tab:mri_parameters}
    \begin{tabular}{@{}lc@{}}
        \toprule
        \multicolumn{1}{c}{Imaging parameter} & Value                   \\ \midrule
        VENC (m/s)                    & $2.5$     \\
        Matrix size                   & $56\times 30\times 113$ \\
        Voxel size (mm$^3$)               & $2\times 2\times 2$     \\
        Oversampling factor           & $2$                     \\
        Cardiac phases                & $30$                    \\
        Time spacing (ms)             & $32.0$                  \\
        $T_2^*$ (ms)  \cite{barthProtonNMRRelaxation1997}                  & $254.0$                 \\
        ADC bandwidth (kHz) \cite{Bender2013}           & $128.0$                   \\ 
        Slew-rate (mT/m/s) \cite{dillingerLimitationsEchoPlanar2020}            & $195.0$                   \\
        Max. gradient amplitude (mT/m) \cite{dillingerLimitationsEchoPlanar2020} & $30.0$              \\ \bottomrule
    \end{tabular}
\end{table}

\subsection{Computation and Comparison of Hemodynamic Parameters}
\label{ssec:hemodynamic_parameters}
In all the experiments conducted in this work, the primary focus was on comparing WSS, OSI, and $\dot{E}_L$ \cite{Barker2014,riva4DFlowEvaluation2021} parameters due to their dependence on blood viscosity. These parameters were estimated as follows:
\begin{subequations}
    \label{eq:hemodynamic-parameters}
    \begin{alignat}{2}
        \text{WSS}(\boldsymbol{x},t) &= \lVert \wss \rVert_2, &&\quad\text{reported in Pascals} \label{eq:subeqwss} \\ 
        \text{OSI}(\boldsymbol{x}) &= \frac{1}{2} \left(1 - \frac{\lVert \int_0^T \wss ~dt \rVert_2}{\int_0^T \lVert \wss \rVert_2 ~dt} \right), &&\quad\text{dimensionless} \label{eq:subeqosi} \\
        \dot{E}_{L}(\boldsymbol{x},t) &= 2\mu\left(\dot{\boldsymbol{\varepsilon}} - \frac{2}{3}\nabla\cdot\boldsymbol{u}\textbf{I}\right):\left(\dot{\boldsymbol{\varepsilon}} - \frac{2}{3}\nabla\cdot\boldsymbol{u}\textbf{I}\right) V(\boldsymbol{x}), &&\quad\text{reported in micro-Watts}  \label{eq:subeqel}
    \end{alignat}
\end{subequations}
Here, $\boldsymbol{l}$ denotes a unit inward normal vector to the surface, $2\mu\dot{\boldsymbol{\varepsilon}}\boldsymbol{l}$ represents the shear stress, $T$ is the duration of the cardiac cycle, $V(\boldsymbol{x})$ is the Voronoi nodal volume distribution on the finite element mesh, and $\textbf{I}$ is the identity matrix. Additionally, $\lVert \cdot \rVert_2$ denotes the standard Euclidean vector norm. In the non-Newtonian case, the apparent viscosity varies both spatially and temporally, and this variation was accounted for by using $\mu_{\text{PL}}$ in the relevant expressions.

These three biomarkers were chosen because of their direct dependence on viscosity. \review{Although near-wall velocity gradients are elevated (suggesting that blood behaves more like a Newtonian fluid in these regions) the pronounced shear-thinning behavior away from the aortic walls (where shear rates are lower), combined with the occurrence of vortex shedding, affects the entire velocity profile and consequently influences the velocity gradients at the wall.} In this context, evaluating WSS is crucial. OSI measures temporal changes in shear stress caused by variations in viscosity and flow dynamics, aspects not captured by $\boldsymbol{t}_{\text{WSS}}$. Finally, $\dot{E}_L$  \review{is a volumetric parameter} quantifying how energy dissipates within the flow due to viscous fluid interactions, which are also sensitive to changes in viscosity, \review{allowing for evaluation away from the vessel walls}.

\begin{figure}[h!]
    \centering
    \includegraphics[width=0.4\textwidth]{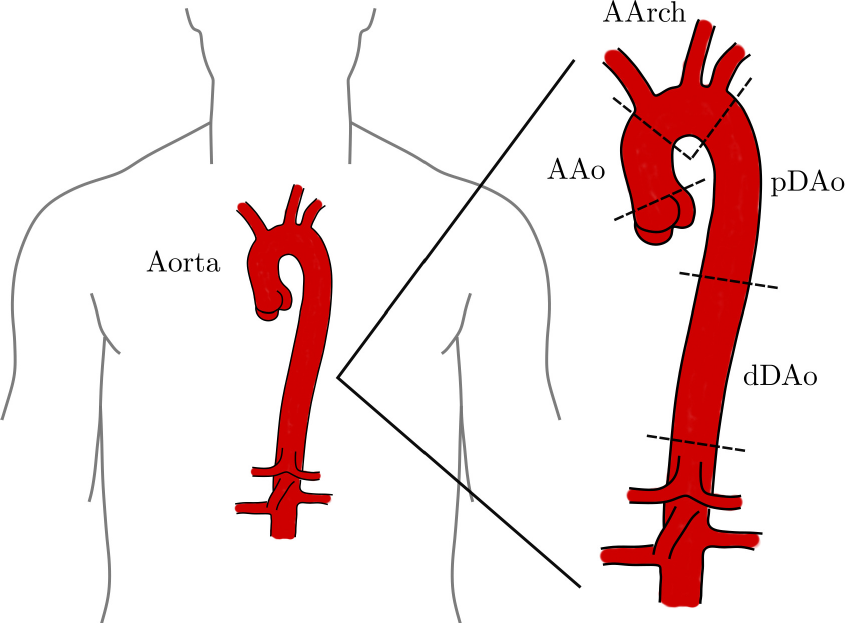}
    \caption{Aortic segments used for the quantification and comparison of hemodynamic parameters.}
    \label{fig:aortic-segments}
\end{figure}

To compare WSS, OSI, and $\dot{E}_L$, the aorta was divided into four segments: the ascending aorta (AAo, between the Valsalva level and the brachiocephalic trunk), the aortic arch (AArch, between the brachiocephalic trunk and the isthmus level), the proximal descending aorta (pDAo, between the isthmus level and the Valsalva level), and the distal descending aorta (dDAo, between the Valsalva level and the diaphragmatic level). This segmentation is not arbitrary and follows the anatomical references used in previous studies \cite{harloff_vivo_2010,schnellKtGRAPPAAccelerated2014,pathroseHighlyAcceleratedAortic2021a,desaiFourDimensionalFlowMagnetic2022,sotelo_fully_2022}. A schematic of these segments is shown in Figure \ref{fig:aortic-segments}. The mean values of WSS, OSI, and $\dot{E}_L$ are then estimated for each segment to facilitate comparisons. In line with current practices \cite{farag_aortic_2018,sotelo3DAxialCircumferential2018,pathroseHighlyAcceleratedAortic2021a,cherryImpact4DFlowMRI2022,ferdianWSSNetAorticWall2022}, the supra-aortic branches (brachiocephalic, carotid, and subclavian arteries) were not included in the analysis due to the small number of voxels relative to artery diameters in this region, which affects the quantification of hemodynamic parameters.

\begin{figure}[h!]
    \centering
    \includegraphics[width=0.65\textwidth]{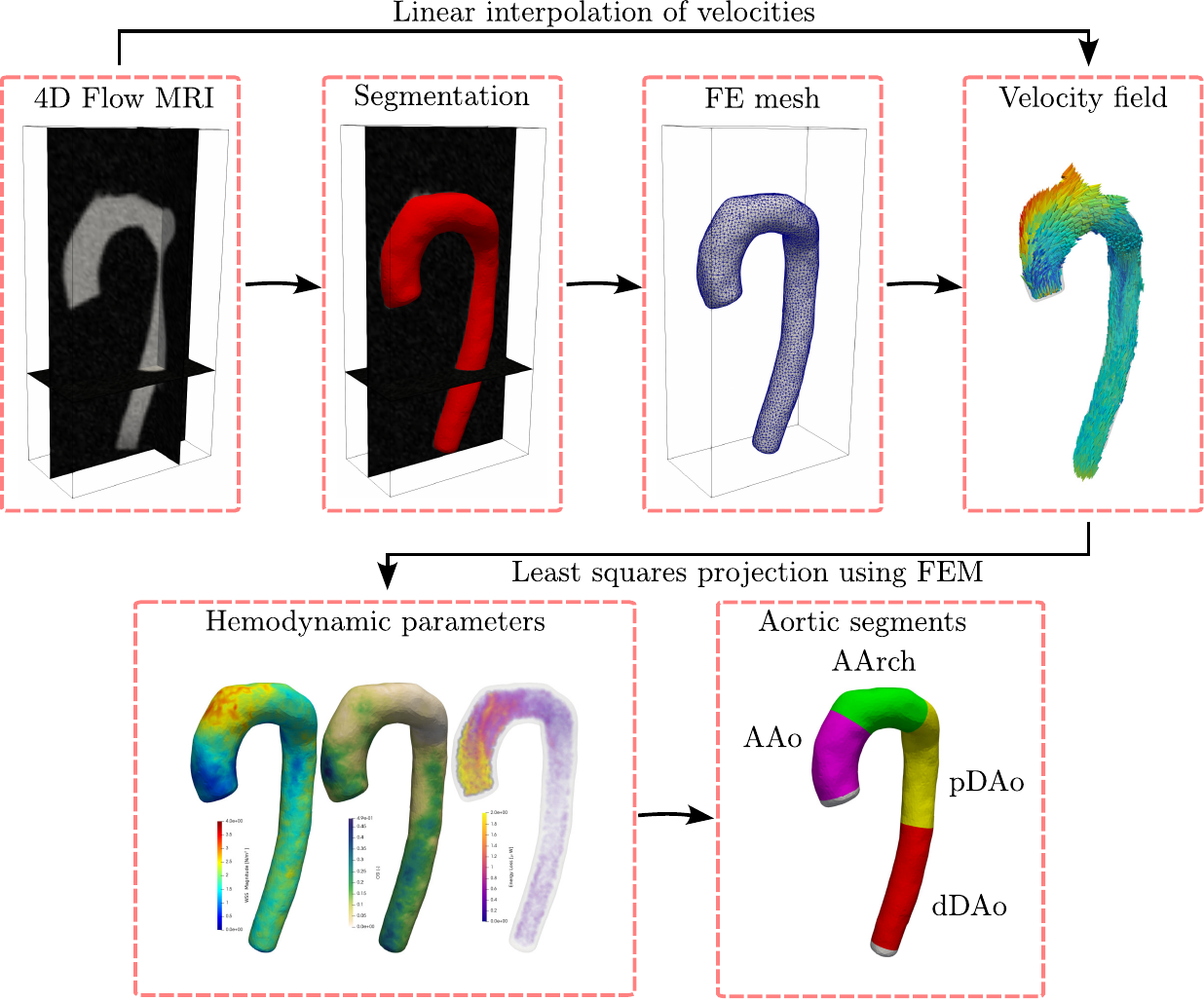}
    \caption{\review{Workflow for estimating hemodynamic parameters from velocity measurements obtained using 4D Flow MRI. First, the image segmentation is derived from the magnitude component of the 4D Flow MRI data. Second, a FE mesh is generated by triangulating the segmented surface and constructing tetrahedral elements for the interior volume. Third, blood velocity data obtained from the image phase information is linearly interpolated onto the FE mesh without any smoothing or correction. Fourth, hemodynamic parameters are estimated by computing velocity gradients through a least-squares projection in the FE sense, as described in \cite{Sotelo2016} and implemented in \cite{Sotelo4dFlowToolbox}. Finally, mean values and standard deviations of the hemodynamic parameters are calculated for each aortic segment.}}
    \label{fig:image_processing}
\end{figure}

The three quantities specified in \eqref{eq:hemodynamic-parameters} were estimated using the 4D Flow Matlab Toolbox \cite{Sotelo4dFlowToolbox}, which employs a previously validated finite element scheme \cite{Sotelo2016} to compute various hemodynamic and geometrical parameters from 4D Flow MRI data \cite{sotelo_fully_2022}. \review{The complete workflow for estimating hemodynamic parameters from 4D Flow MR images is shown in Figure \ref{fig:image_processing}.}

\subsection{Experiment 1: evaluating the pertinence of the power-law}
\label{ssec:experiment_1}
This expermient aims to assess using CFD simulations whether Newtonian and non-Newtonian viscosity models yield equivalent results in terms of WSS, $\dot{E}_L$, and OSI across different Hct levels. The goal is to determine if a constant viscosity model suffices to account for the effects of nonlinear viscosity on hemodynamic parameters. If the results between the models are similar, a constant viscosity would be deemed adequate, \review{supporting the current MRI community's practice of using a constant viscosity for estimating hemodynamic parameters from 4D Flow measurements}. Otherwise, the adoption of a more complex model would be necessary.

To evaluate this, CFD simulations of a patient-specific aorta model (see Figure \ref{fig:workflow}) were performed for the Hct values listed in Table \ref{tab:viscosities}, using both the Newtonian fitted viscosities ($\mu_{\text{NF}}$) and power-law viscosities ($\mu_{\text{PL}}$) estimated as described in Section \ref{ssec:newtonian visco estimation}. Details of the model are provided in Section \ref{ssec:flow simulation}. The resulting velocity fields were then used to generate synthetic 4D Flow MR images as described in Section \ref{ssec:image generation}, which were \review{image}-processed to obtain WSS, $\dot{E}_L$, and OSI using both $\mu_{\text{NF}}$ and $\mu_{\text{PL}}$ \review{with the software and procedure describe in Section \ref{ssec:hemodynamic_parameters}}.

To facilitate comparisons, mean and standard deviation values of all hemodynamic parameters within each segment, as described in Section \ref{ssec:hemodynamic_parameters}, were computed at peak systole and diastole (see Figure \ref{fig:flows_segments}). The reported values represent the average and standard deviation of the means across the four segments.

\subsection{Experiment 2: comparison of current approaches with power-Law model parameter estimation}
\label{ssec:experiment_2}
\review{Since blood exhibits shear-thinning behavior, this experiment aims to highlight the differences in hemodynamic parameter estimation when using a Newtonian viscosity ($\mu_{\text{N}}$) versus a shear-thinning model ($\mu_{\text{PL}}$). To do so, we used the same 4D Flow images generated in Experiment 1 from power-law viscosity CFD simulations to compute WSS, $\dot{E}_L$, and OSI at different Hct levels. These parameters were estimated using both Newtonian (within the typical range found in the literature) and power-law viscosities as described in Section \ref{ssec:newtonian visco estimation}. Notably, the hemodynamic parameters estimated using $\mu_{\text{PL}}$ are identical to those from Experiment 1.}

\begin{figure}[!htb]
    \centering
    \includegraphics[width=0.7\textwidth]{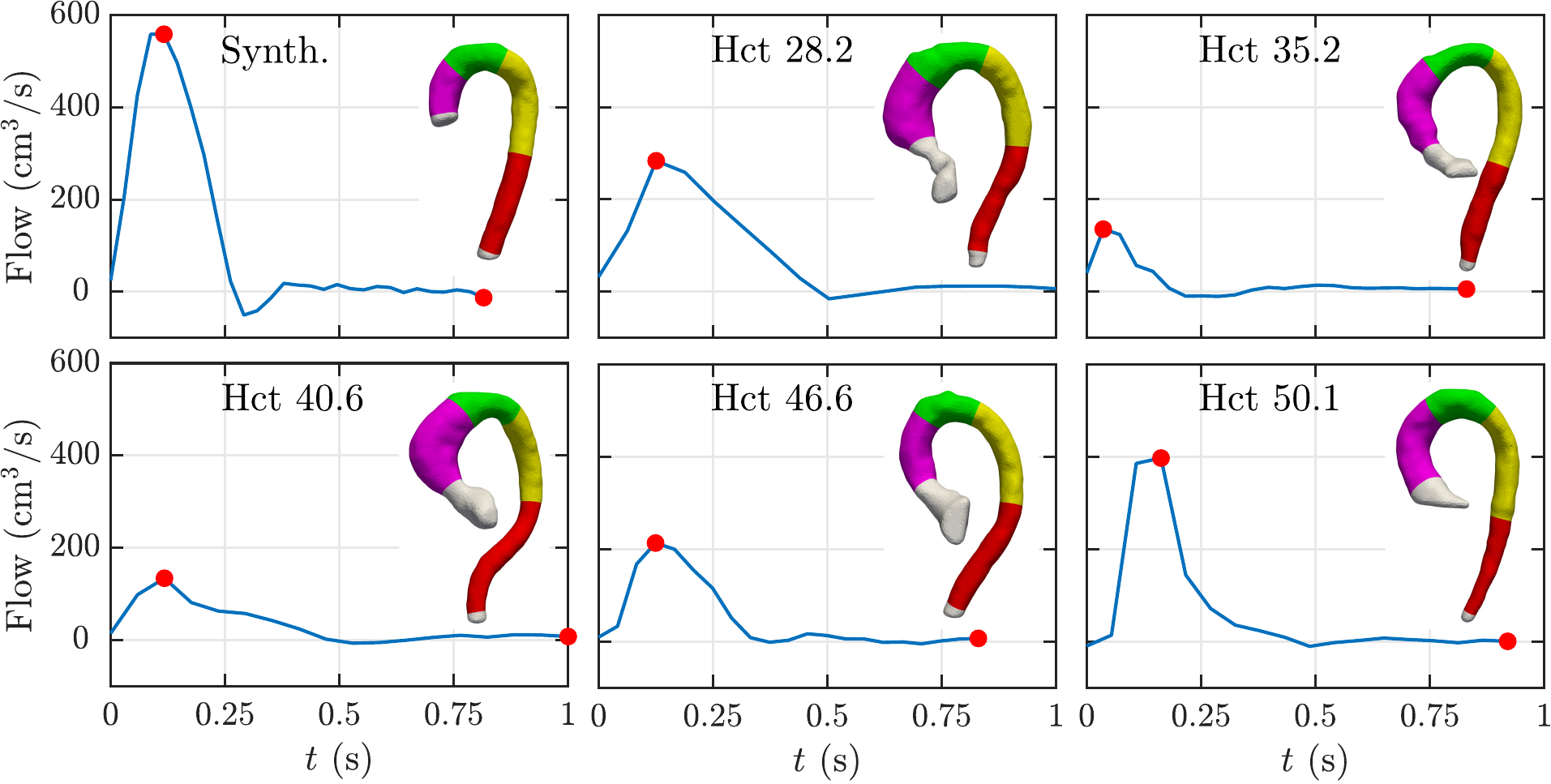} 
    \caption{Flow curves obtained from 4D Flow images and the aortic segments considered in all analyses. The flow curves were measured at the ascending aorta, at the level of the pulmonary bifurcation. The top-left image shows the measurements from the synthetic dataset, while the remaining figures display the curves obtained from patient data.}
    \label{fig:flows_segments}
\end{figure}

Once again, to facilitate comparisons, mean and standard deviation values of all hemodynamic parameters within each segment, as described in Section \ref{ssec:hemodynamic_parameters}, were computed at peak systole and diastole (see Figure \ref{fig:flows_segments}). The reported values represent the average and standard deviation of the means across the four segments.

\subsection{Experiment 3: in-vivo data}
\label{ssec:experiment_3}
This experiment aims to replicate the results obtained in Experiment 2 using in-vivo data from patients with varying Hct levels, following the same processing procedure described previously. For this purpose, in-vivo gadolinium-enhanced triple-VENC 4D Flow images acquired from 5 patients with hypertrophic cardiomyopathy (2 males and 3 females, aged $68 \pm 26$ years) using a 3.0T Achieva scanner (Philips Healthcare, Best, The Netherlands) were used. These patient data are part of a larger dataset that has been previously utilized in \cite{iwataMeasurementTurbulentKinetic2024} for the quantification of turbulent kinetic energy. The criterion used to choose the data was to have equispaced samples within the broadest Hct range possible. The imaging parameters included an acquired voxel size of 1.7x1.7x2.0 mm$^3$, VENCs of 50, 150, and 450 cm/s, a temporal resolution of 40 ms, and 15 to 21 frames depending on heart rate. Further details of the sequence and reconstruction can be found in \cite{iwataMeasurementTurbulentKinetic2024}.

Each patient had a Complete Blood Count (CBC) available, which is a comprehensive blood test that includes Hct measurement. Due to the use of gadolinium as a contrast agent, a kidney function test was required to determine its suitability. Therefore, blood tests were mandated within three months prior to imaging, which also included a complete blood count. The patients had Hct levels of 28.2, 35.2, 40.2, 46.6, and 50.1.

The experiments were conducted at Nippon Medical School Hospital, with the study approved by the Institutional Review Board. All subjects provided written informed consent.

\section{Results}
\label{sec:results}
As described in Section \ref{ssec:hemodynamic_parameters}, only the mean values across the four segments for each hemodynamic parameter, along with their respective standard deviations (representing variations across segments), are shown in this section to avoid visual clutter. Figures with detailed regional information for each segment are presented in Supplementary Figures S2 to S4.

\begin{figure*}[!hbt]
    \centering
    \includegraphics[width=\textwidth]{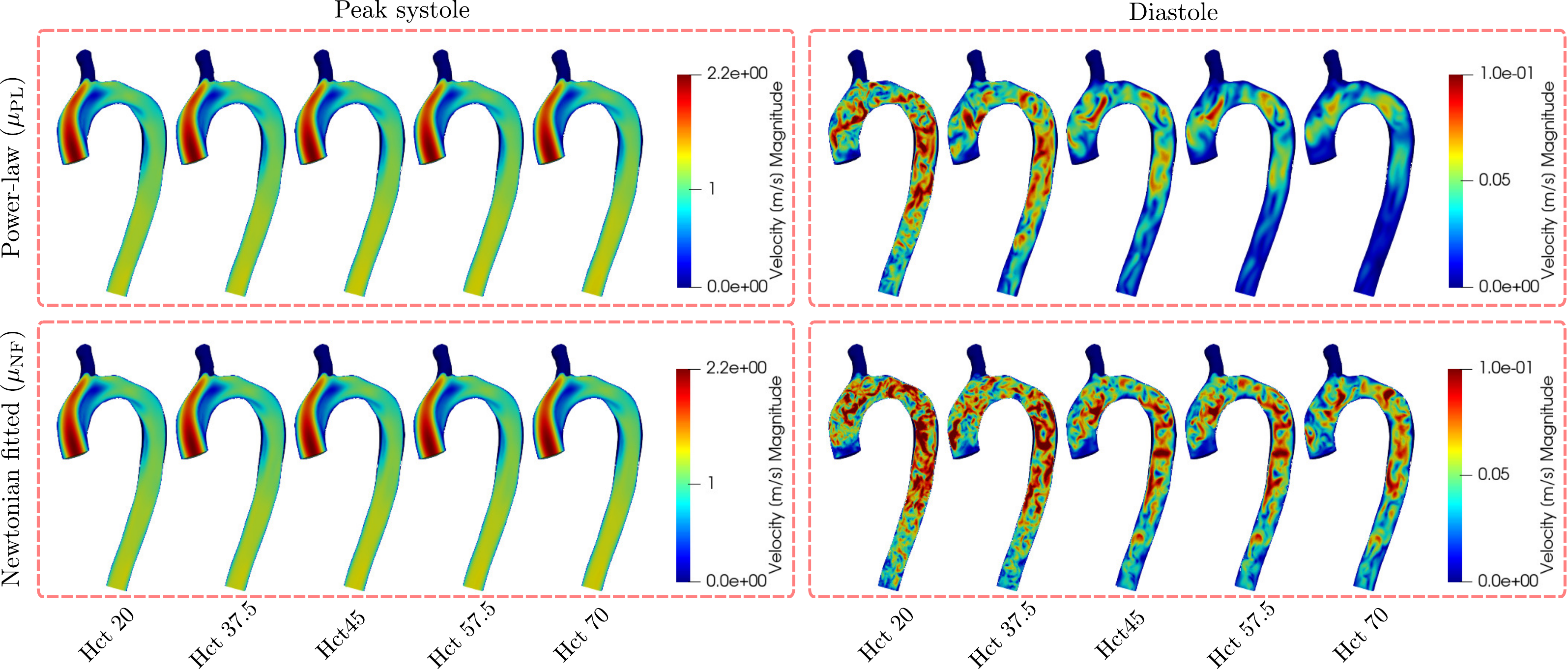}
    \caption{Velocity maps at peak systole and diastole, extracted from CFD simulations using power-law and Newtonian fitted viscosities for different Hct values. Almost identical flow patterns are observed at peak systole, while great differences are noticeable at diastole.}
    \label{fig:simulation velocities}
\end{figure*}

\subsection{Experiment 1}
\label{sec:results experiment 1}
\begin{figure}[!htb]
    \centering
    \includegraphics[width=0.6\textwidth]{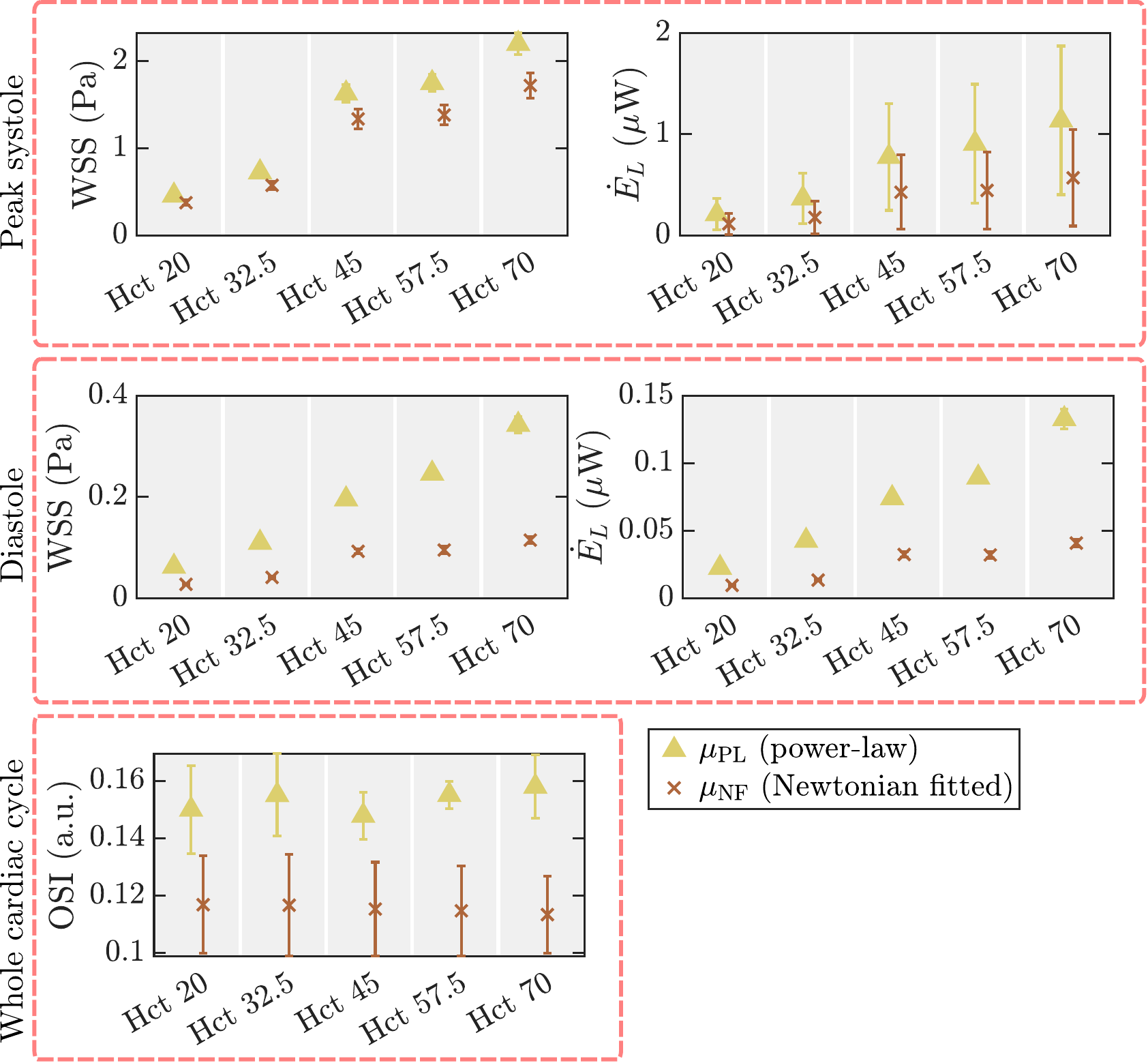}
    \caption{Mean values and standard deviations of WSS, OSI, and $\dot{E}_L$ across four aortic segments were estimated from simulated images obtained from CFD simulations using power-law ($\mu_{\mathrm{PL}}$) and Newtonian fitted ($\mu_{\mathrm{NF}}$) viscosities. The estimations were performed using viscosity models and values consistent with those used in the CFD simulations from which synthetic 4D Flow images were obtained. Clear differences were observed between the models for varying Hct values and cardiac phases, indicating that the two models cannot be used interchangeably.}
    \label{fig:hem_params_adj_sims}
\end{figure}

Figure \ref{fig:simulation velocities} shows the simulated velocities obtained using power-law ($\mu_{\mathrm{PL}}$) and Newtonian fitted ($\mu_{\mathrm{NF}}$) viscosities. \review{Both models produced nearly identical velocity profiles at peak systole across all Hct levels, while notable differences were observed during diastole. For both models, as the Hct increases (and consequently the viscosity), fluid vorticity decreases at diastole, with this effect being more pronounced in the power-law simulations.}

\review{Despite the almost identical velocity maps in systole, Figure \ref{fig:hem_params_adj_sims} shows that hemodynamic parameters estimated from synthetic 4D Flow images exhibit noticeable differences at both systole and diastole}. At systole, the average differences in WSS and $\dot{E}_L$ across all Hct levels, relative to the power-law model, were $-22.0\pm 1.8$\% and $-44.7\pm 3.8$\%, respectively. In absolute terms, these differences range from $-0.09$ to $-0.52$ Pa in WSS and from $-0.08$ to $-0.54$ $\mu$W in $\dot{E}_L$ for Hct values of 20 and 70, respectively. In all cases, the sign indicates whether the parameter estimated using Newtonian fitted ($\mu_{\mathrm{NF}}$) viscosities was greater (positive) or smaller (negative) compared to the power-law estimations.

In diastole, \review{the relative differences become more pronounced}. The average differences in WSS and $\dot{E}_L$ across all Hct levels were $-59.9\pm 5.5$\% and $-63.3\pm 5.9$\%, respectively. In absolute terms, these differences range from $-0.04$ to $-0.23$ Pa in WSS and from $-0.01$ to $-0.09$ $\mu$W in $\dot{E}_L$ for Hct values of 20 and 70, respectively. Although the absolute differences are smaller, the relative differences suggest that using a Newtonian model for blood, even when adjusted by Hct level, is not suitable for capturing the nonlinear interactions of blood flow. Consequently, a Newtonian model cannot serve as a suitable surrogate, reinforcing the need to incorporate nonlinear rheology \review{to the estimation of hemodynamic parameters from velocity measurements obtained using 4D Flow MRI.}

For OSI, the average difference across all Hct levels was $-24.6 \pm 2.6$\%. In absolute terms, the differences ranged from $-0.03$ to $-0.05$ for Hct values between 20 and 70, respectively. Since OSI is a metric calculated over the entire cardiac cycle, the negative sign indicates that the WSS vector experienced greater variation when the blood exhibited nonlinear behavior, aligning with our previous findings.

Additionally, greater variability between segments, \review{as indicated by the standard deviation bars}, was observed in $\dot{E}_L$ at peak systole and OSI compared to WSS (see Figure \ref{fig:hem_params_adj_sims}). This is further illustrated in Figure \ref{fig:appendix-model-interrogation} in \ref{sec:sample:appendix}, which presents the mean and standard deviation of hemodynamic parameters within each aortic segment. For $\dot{E}_L$, higher values were observed in the ascending aorta, \review{with a decreasing trend along the artery}, primarily due to the flow dynamics (see Figure \ref{fig:simulation velocities}). \review{A similar pattern is observed for OSI, although the trend is less pronounced in the graph.}

\begin{figure*}[!hbt]
    \centering
    \includegraphics[width=\textwidth]{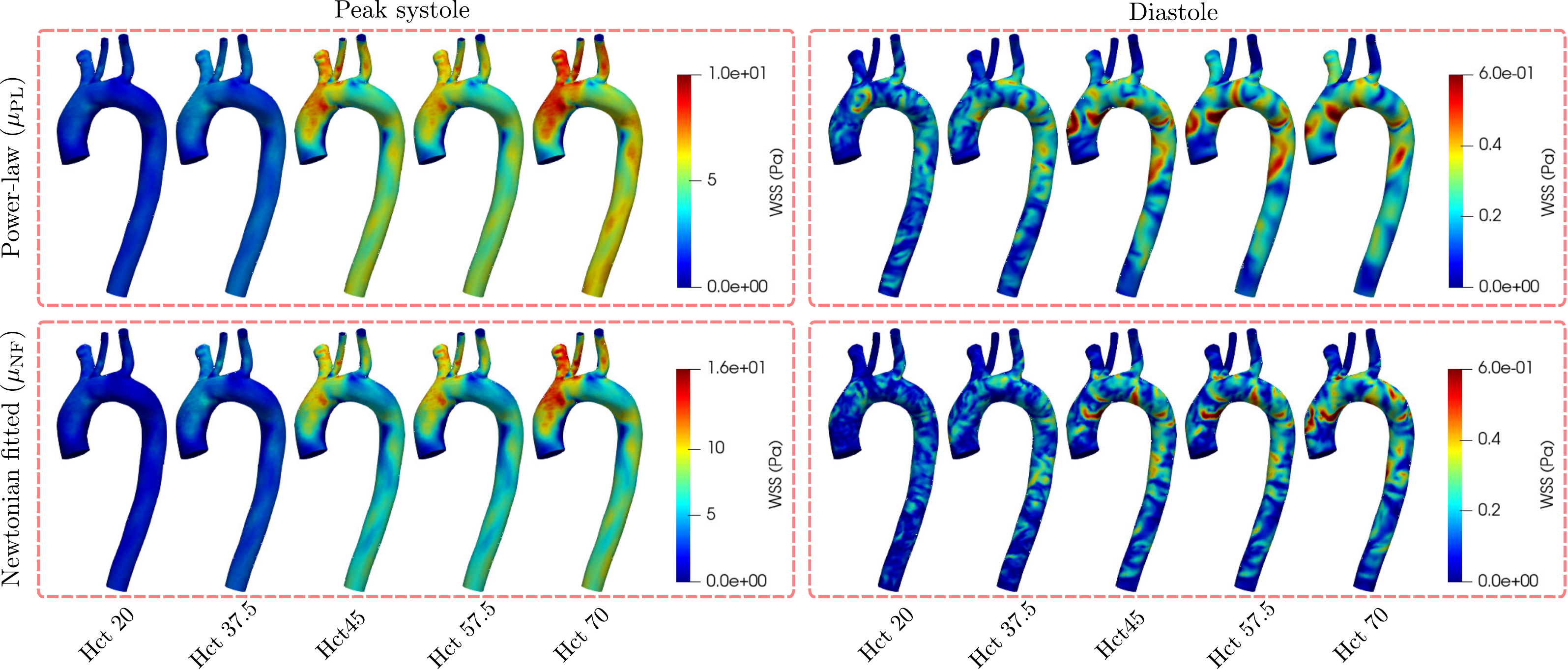}
    \caption{WSS maps at peak systole and diastole were estimated from CFD simulations using power-law ($\mu_{\mathrm{PL}}$) and Newtonian fitted ($\mu_{\mathrm{NF}}$) viscosities for different Hct values. \review{The color scale for Newtonian fitted results is higher at peak systole}. The WSS obtained from the Newtonian fitted simulations is greater at peak systole but smaller at diastole compared to those obtained using the power-law model, indicating that the two simulations cannot be used interchangeably.}
    \label{fig:wss_comparison_simulations}
\end{figure*}

Finally, Figure \ref{fig:wss_comparison_simulations} presents the WSS calculated directly from the FE simulations using both models at peak systole and diastole. \review{While differences between models are clearly observable in both MR images and CFD simulations, a notable discrepancy in behavior was observed at peak systole when analyzing the WSS obtained from CFD simulations. In MRI results, WSS estimated using the power-law viscosity model was higher than that obtained with its Newtonian counterpart (see Figure \ref{fig:hem_params_adj_sims}). However, in CFD results, the opposite trend was observed, with the Newtonian model yielding higher WSS values than the power-law model. This finding underscores the importance of incorporating a non-Newtonian model, as WSS estimations using the power-law model in MRI exhibited less underestimation compared to its Newtonian counterpart against CFD results.}

\subsection{Experiment 2}
\label{sec:results experiment 2}
\begin{figure}[!htb]
    \centering
    \includegraphics[width=0.6\textwidth]{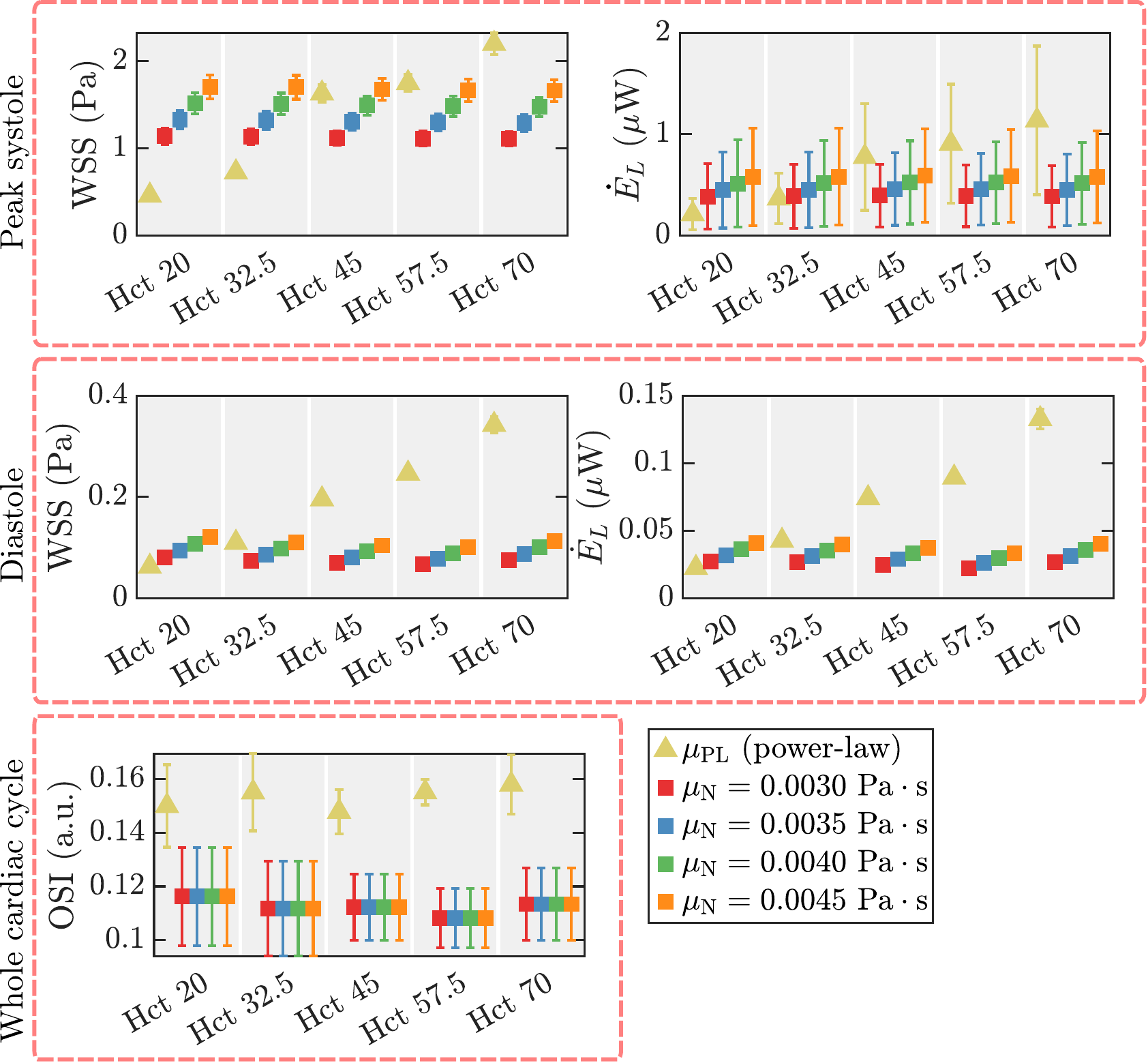}
    \caption{Mean values and standard deviations of WSS, OSI, and $\dot{E}_L$ across four aortic segments were estimated from synthetic 4D Flow images generated by CFD simulations using power-law viscosities. Hemodynamic parameters were estimated using both power-law ($\mu_{\mathrm{PL}}$) and state-of-the-art Newtonian ($\mu_{\mathrm{N}}$) viscosities. Clear differences were observed between the models across varying Hct values and cardiac phases, indicating that the choice of viscosity model can lead to significant discrepancies in hemodynamic parameter estimations.}
    \label{fig:hem_params_sims}
\end{figure}

\begin{figure}[!hbt]
    \centering
    \includegraphics[width=0.5\textwidth]{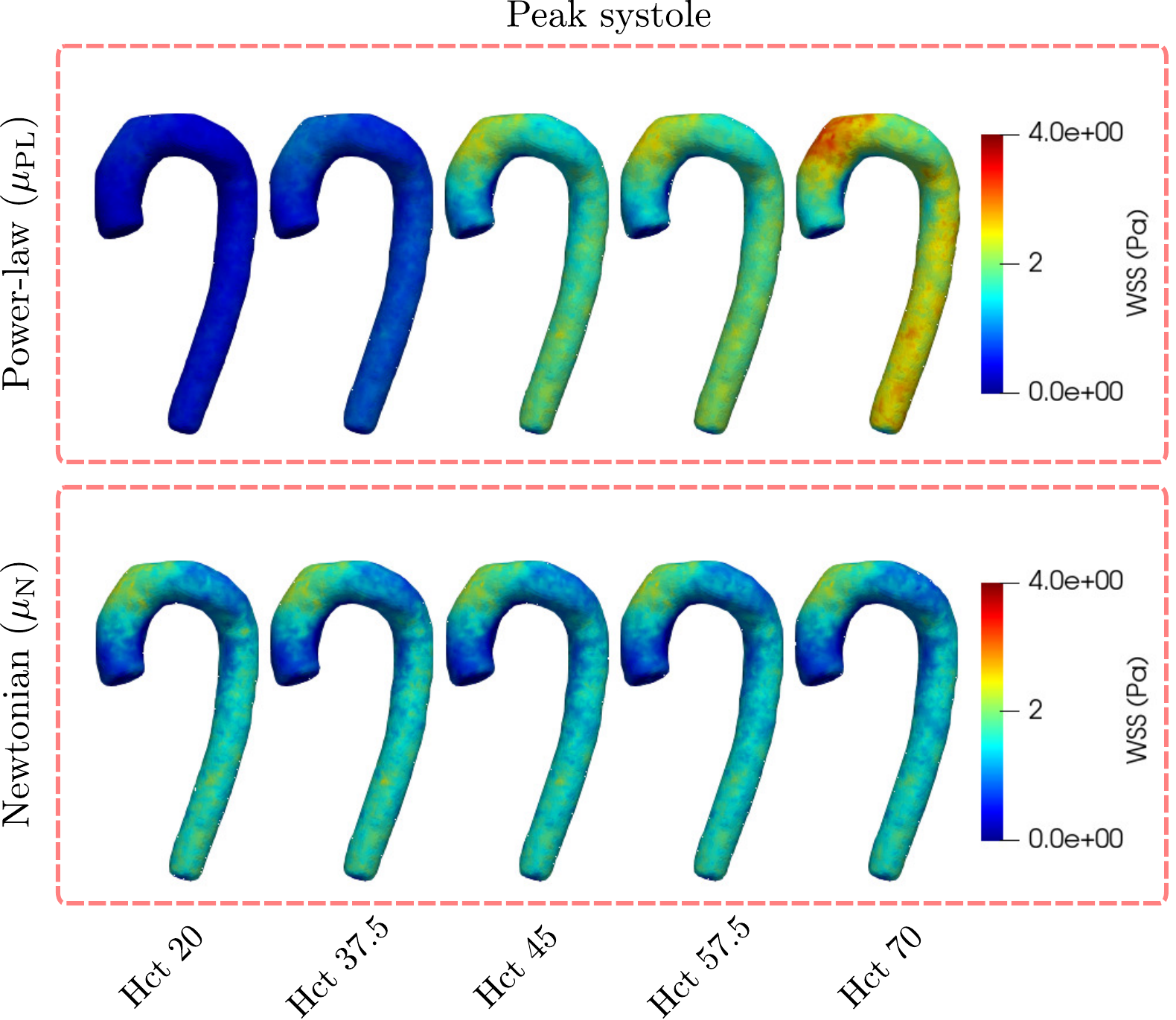}
    \caption{WSS maps at peak systole estimated from synthetic 4D Flow MR images obtained from power-law CFD simulations. The estimations were obtained using power-law ($\mu_{\mathrm{PL}}$) viscosities (top row) and a Newtonian viscosity of $\mu_{\mathrm{N}} = 0.0035$ $\text{Pa}\cdot\text{s}$ (bottom row).}
    \label{fig:wss_maps_sims_sys}
\end{figure}

Given that blood exhibits shear-thinning behavior \review{(we emphasize that our study builds upon this)}, Figure \ref{fig:hem_params_sims} shows the hemodynamic parameters estimated from synthetic 4D Flow images generated from simulations with a power-law blood rheology. The parameters were estimated using both power-law ($\mu_{\mathrm{PL}}$) and standard Newtonian ($\mu_{\mathrm{N}}$) viscosities commonly found in the literature. At systole, the differences in WSS and $\dot{E}_L$ across all Hct levels (relative to the power-law model and using $\mu_{\mathrm{N}}=0.0035~\text{Pa}\cdot\text{s}$ as reference) ranged from $+189.4\%$ to $-41.6\%$ and from $+112.1\%$ to $-60.4\%$ for Hct values of 20 and 70, respectively. In absolute terms, these differences ranged from $+0.87$ to $-0.92$ Pa in WSS and from $+0.23$ to $-0.68$ $\mu$W in $\dot{E}_L$ for Hct values from 20 to 70, respectively. The sign change indicates that WSS and $\dot{E}_L$ estimated using Newtonian viscosities can be either greater (positive) or smaller (negative) than the power-law estimations, depending on the Hct level. These differences can be better observed in Figure \ref{fig:wss_maps_sims_sys} for WSS maps obtained on the segmented aorta at peak systole.

\begin{figure}[!hbt]
    \centering
    \includegraphics[width=0.5\textwidth]{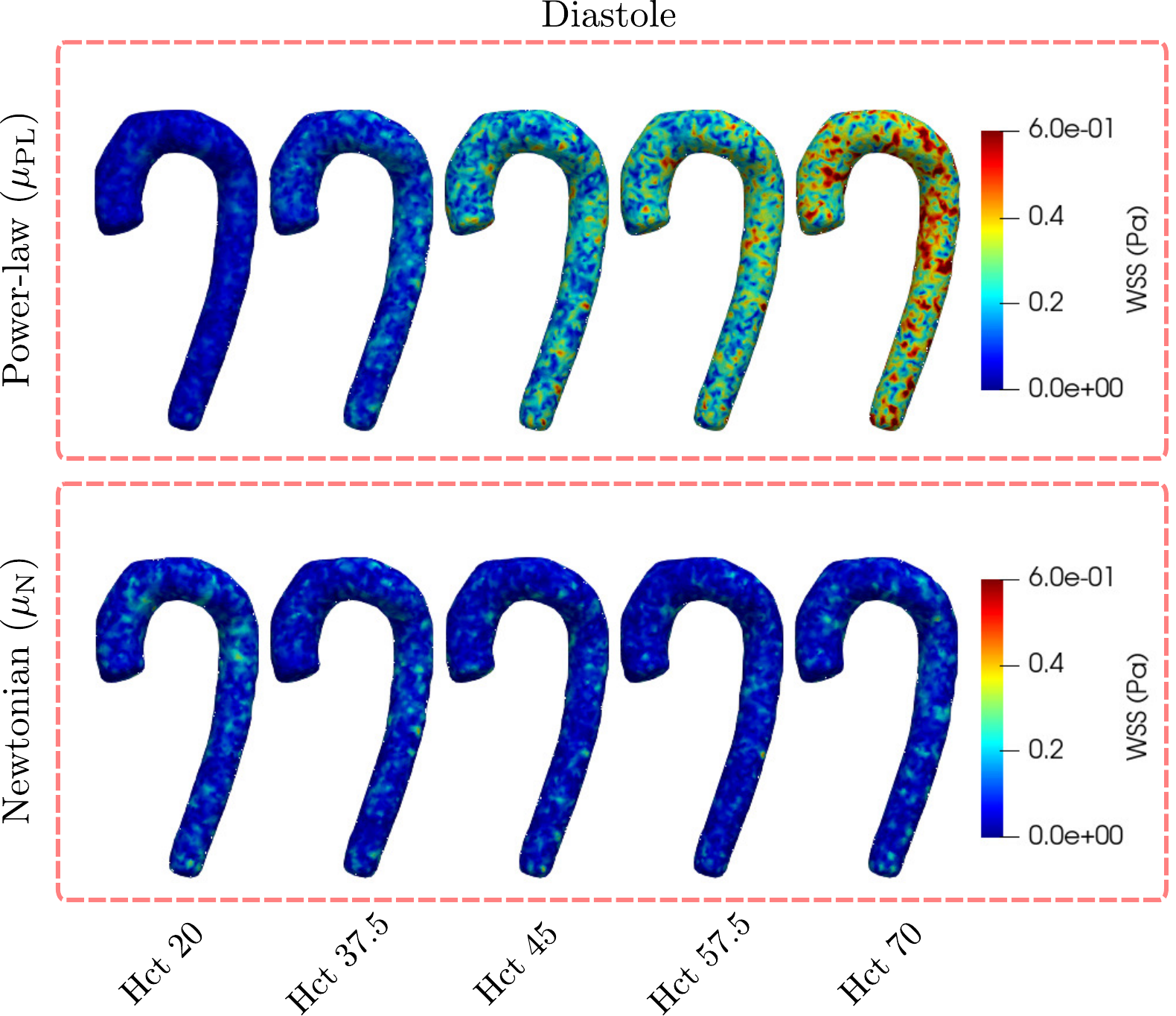}
    \caption{WSS maps at diastole estimated from synthetic 4D Flow MR images obtained from power-law CFD simulations. The estimations were obtained using power-law ($\mu_{\mathrm{PL}}$) viscosities (top row) and a Newtonian viscosity of $\mu_{\mathrm{N}} = 0.0035$ $\text{Pa}\cdot\text{s}$ (bottom row).}
    \label{fig:wss_maps_sims_dias}
\end{figure}

In diastole, the results were similar, with relative differences in WSS ranging from $+51.5\%$ to $-74.4\%$ (using $\mu_{\mathrm{N}}=0.0035~\text{Pa}\cdot\text{s}$ as reference) and from $+20.3\%$ to $-79.9\%$ in $\dot{E}_L$. In absolute terms, these differences ranged from $+0.03$ to $-0.26$ Pa in WSS and from $+0.01$ to $-0.10$ $\mu$W in $\dot{E}_L$ for Hct values from 20 to 70, respectively. \review{In contrast to the systolic results, both WSS and $\dot{E}_L$ estimated at diastole using the power-law viscosity model were higher than those obtained with Newtonian viscosities at nearly all Hct levels.} These differences are further illustrated in Figure \ref{fig:wss_maps_sims_dias}, which presents WSS maps of the segmented aorta obtained from 4D Flow MR images at diastole.

\begin{figure*}[!hbt]
    \centering
    \includegraphics[width=\textwidth]{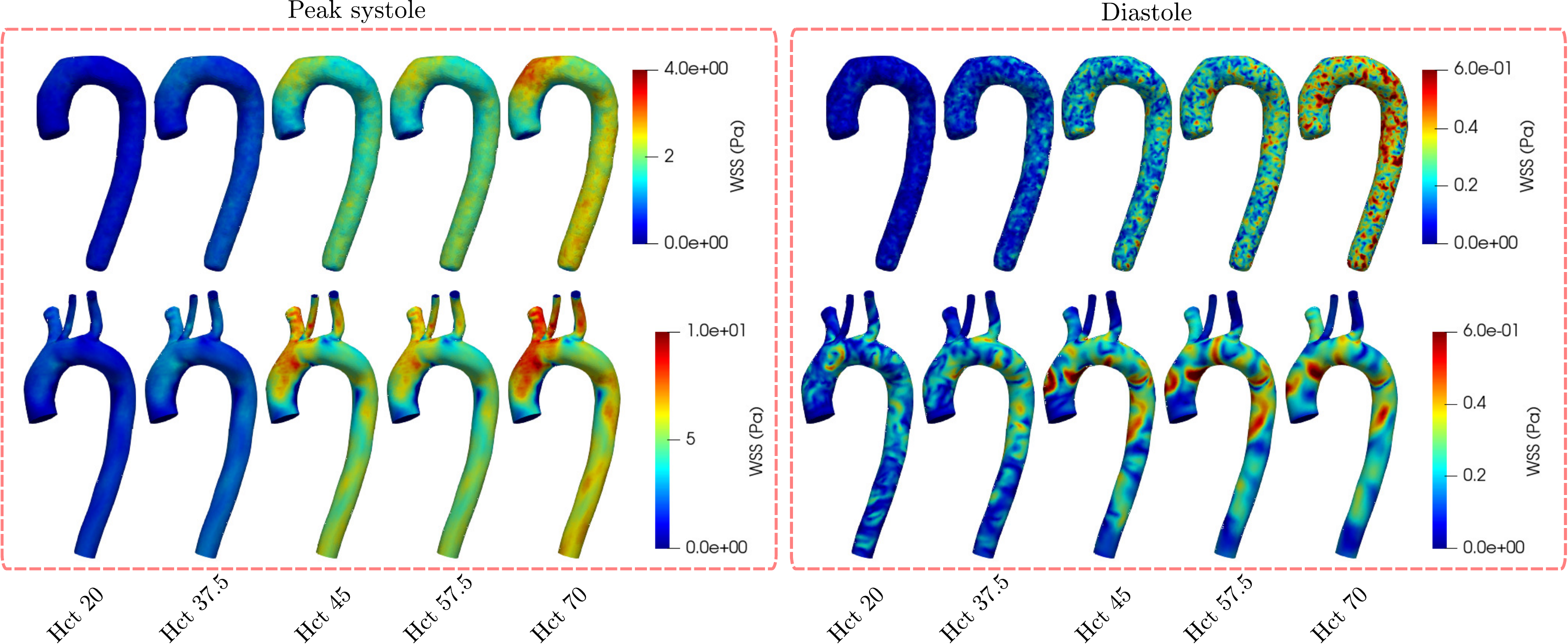}
    \caption{WSS maps at peak systole and diastole estimated from synthetic 4D Flow images generated by CFD simulations using power-law viscosities. The color bars at peak systole have different ranges to clearly display the spatial distributions. The results reveal similar spatial variations in WSS between the images and simulations, despite the presence of noise, partial volume effects, and flow artifacts, at both peak systole and diastole. However, consistent with previous reports in the literature, WSS values estimated from 4D Flow images tend to underestimate those obtained from simulations.}
\end{figure*}

For OSI, using $\mu_{\mathrm{N}}=0.0035~\text{Pa}\cdot\text{s}$ as the reference, the relative differences with respect to the power-law estimations ranged from $-22.5\%$ to $-30.2\%$ for Hct values between 20 and 70. In absolute terms, the differences ranged from $-0.03$ to $-0.05$ over the same Hct range. As before, the negative sign indicates that the WSS vector exhibited greater temporal variation when blood was modeled with nonlinear behavior.

Similar to the results in Experiment 1, greater variability between segments was observed in $\dot{E}_L$ at peak systole and OSI compared to WSS (see Figure \ref{fig:hem_params_sims}). This is more clearly illustrated in Figure \ref{fig:appendix-2} in \ref{sec:sample:appendix}, which shows the means and standard deviations of hemodynamic parameters within each aortic segment.

Since we have demonstrated differences at peak systole and diastole between estimations made using both viscosity models ($\mu_{\mathrm{N}}$ and $\mu_{\mathrm{PL}}$), it is important to assess whether the parameters estimated using the power-law model are consistent with the simulations. These similarities are illustrated in Figure \ref{fig:wss_comparison_simulations}, which shows that the WSS maps estimated from MR images and CFD simulations exhibit comparable spatial variations. However, lower WSS values were observed at peak systole in synthetic 4D Flow MR images compared to their simulation counterparts, with differences reaching approximately 60\% of the CFD simulation values. At diastole, the WSS values were more comparable, although the estimations obtained from MR images were notably affected by noise.

\subsection{Experiment 3}
\label{sec:results experiment 3}
\begin{figure}[!hbt]
    \centering
    \includegraphics[width=0.6\textwidth]{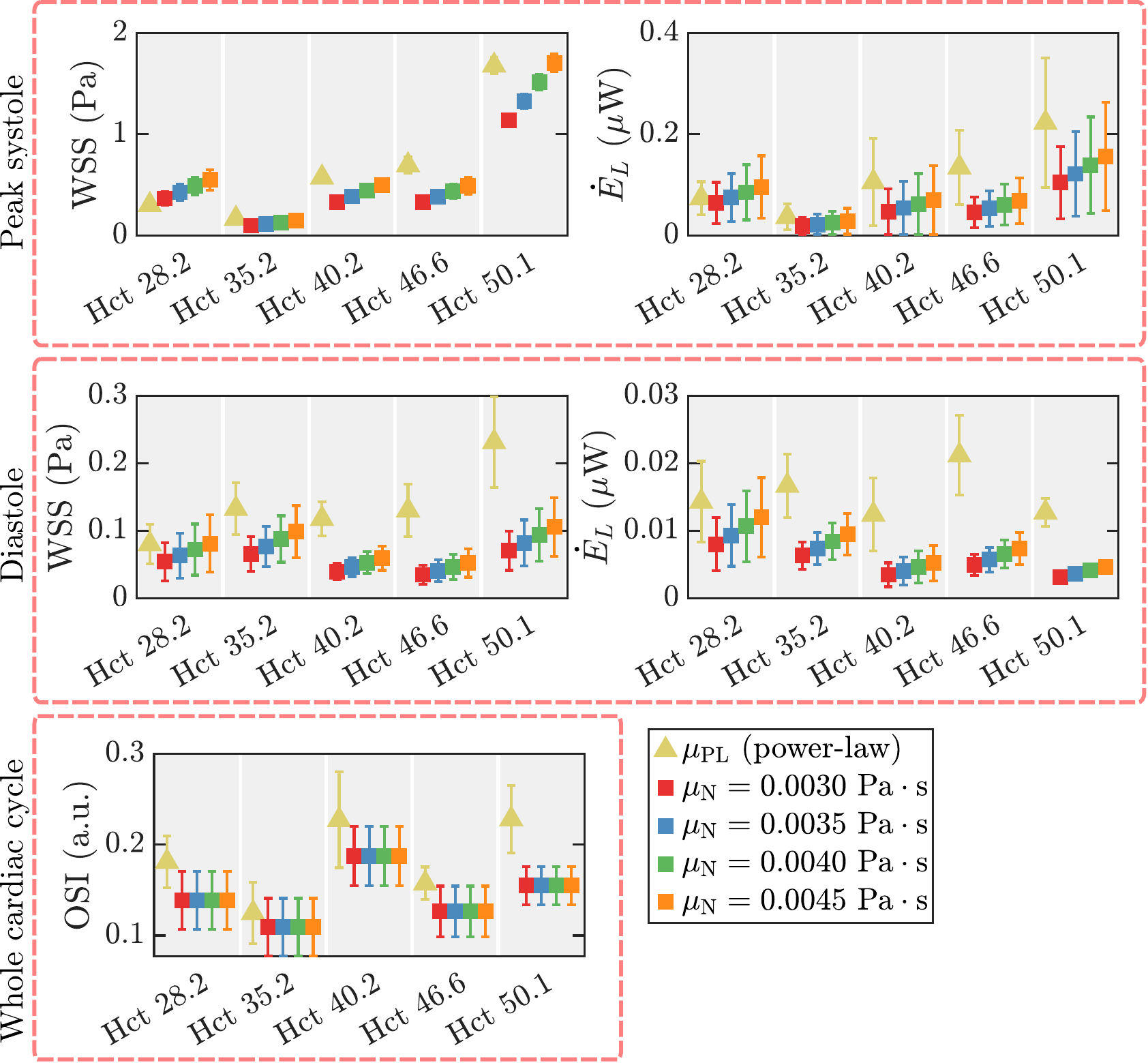}
    \caption{Mean values and standard deviations of WSS, OSI, and $\dot{E}L$ across four aortic segments were estimated from in-vivo 4D Flow images of HCM patients. Hemodynamic parameters were calculated using both power-law ($\mu{\mathrm{PL}}$) and state-of-the-art Newtonian ($\mu_{\mathrm{N}}$) viscosities. Clear differences were observed between the models across varying Hct values and cardiac phases, highlighting that the choice of viscosity model can result in significant discrepancies in hemodynamic parameter estimations.}
    \label{fig:hem_params_invivo}
\end{figure}

Figure \ref{fig:hem_params_invivo} presents the hemodynamic parameters estimated from in-vivo 4D Flow MR images. It is important to note that the Hct range in this experiment differs from that used in Experiments 1 and 2, ranging from 28.2 to 50.1, compared to 20 to 70 in the earlier experiments. Additionally, the Hct values in this case are not equidistant. As shown in the figure, estimating hemodynamic parameters using the power-law viscosity model ($\mu_{\mathrm{PL}}$) generally resulted in higher values across nearly all cases. The only exception occurred at an Hct of 28.2 during peak systole, where WSS and $\dot{E}L$ values obtained with $\mu{\mathrm{PL}}$ were lower.

At peak systole, for an Hct of 28.2 (the lowest), the relative differences with respect to the power-law estimation (using $\mu_{\mathrm{N}}=0.0035~\text{Pa}\cdot\text{s}$ as a reference) were $+41.4\%$ in WSS and $+1.5\%$ in $\dot{E}_L$. For an Hct of 50.1 (the highest), the differences were $-21.0\%$ and $-45.4\%$, respectively. However, the largest differences were observed at an Hct of 46.6, with $-45.0\%$ in WSS and $-60.4\%$ in $\dot{E}_L$. In absolute terms, these differences in WSS and $\dot{E}_L$ were $+0.125$ Pa and $+0.001$ $\mu$W, $-0.543$ Pa and $-0.101$ $\mu$W, and $-0.312$ Pa and $-0.081$ $\mu$W for Hct levels of 28.2, 50.1, and 46.6, respectively.

In diastole, these differences become larger. For an Hct of 28.2, the relative differences were $-21.4\%$ in WSS and $-34.8\%$ in $\dot{E}_L$, while for an Hct of 50.1, they were $-64.6\%$ and $-71.8\%$, respectively. However, as with the peak-systolic results, the largest differences were observed at an Hct of 46.6, with $-69.0\%$ in WSS and $+73.0\%$ in $\dot{E}_L$. In absolute terms, these differences in WSS and $\dot{E}_L$ were $-0.017$ Pa and $-0.005$ $\mu$W, $-0.149$ Pa and $-0.009$ $\mu$W, and $-0.090$ Pa and $-0.016$ $\mu$W for Hct levels of 28.2, 50.1, and 46.6, respectively.

For OSI, the relative differences with respect to the power-law estimations averaged $-21.1 \pm 6.5\%$ across all Hct values considered in the experiment. In absolute terms, this variation averaged $-0.041 \pm 0.019$. These results indicate that using $\mu_{\mathrm{PL}}$ to estimate WSS reveals greater temporal variations, consistent with the findings from Experiment 2.

Although smaller differences between viscosity models were observed in the in-vivo MR data (particularly at peak systole), these discrepancies can be attributed to the differing blood flow dynamics and lower flow rates found in patients compared to the CFD simulations (see Figure \ref{fig:flows_segments}). The reduced flow likely results in smaller velocity gradients near the vessel walls, leading to lower estimations of WSS and $\dot{E}_L$. OSI, however, is a parameter normalized over time, which helps preserve the observed differences between models. \review{Additionally, these smaller differences are accentuated by the narrower Hct range considered in this experiment.}

\begin{figure}[!hbt]
    \centering
    \includegraphics[width=0.6\textwidth]{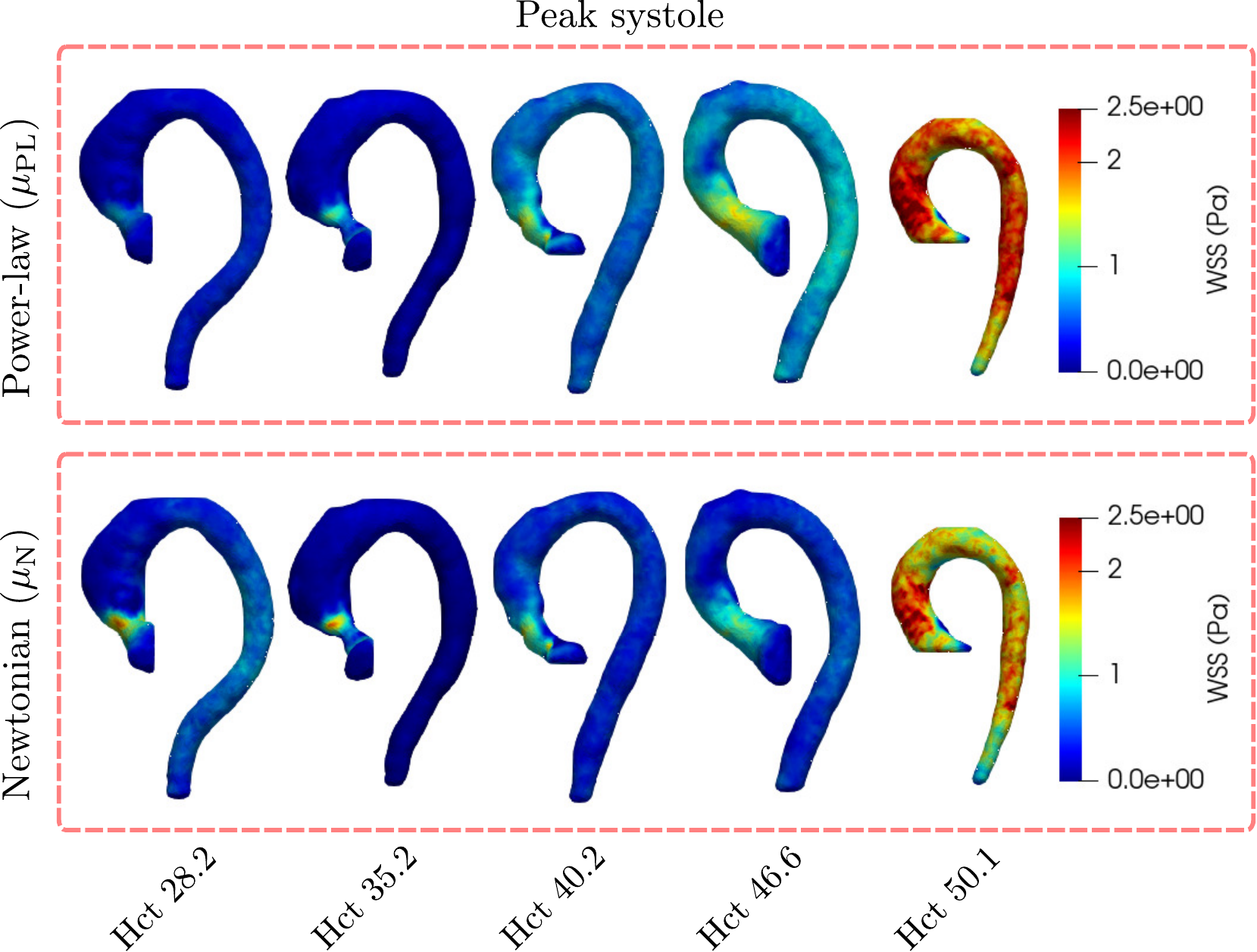}
    \caption{WSS maps at peak systole estimated from in-vivo 4D Flow images of patients with HCM at different Hct levels. The estimations shown were obtained using power-law ($\mu_{\mathrm{PL}}$) viscosities and a Newtonian viscosity of $\mu_{\mathrm{N}} = 0.0035$ $\text{Pa}\cdot\text{s}$. }
    \label{fig:wss_maps_invivo_sys}
\end{figure}

\begin{figure}[!hbt]
    \centering
    \includegraphics[width=0.6\textwidth]{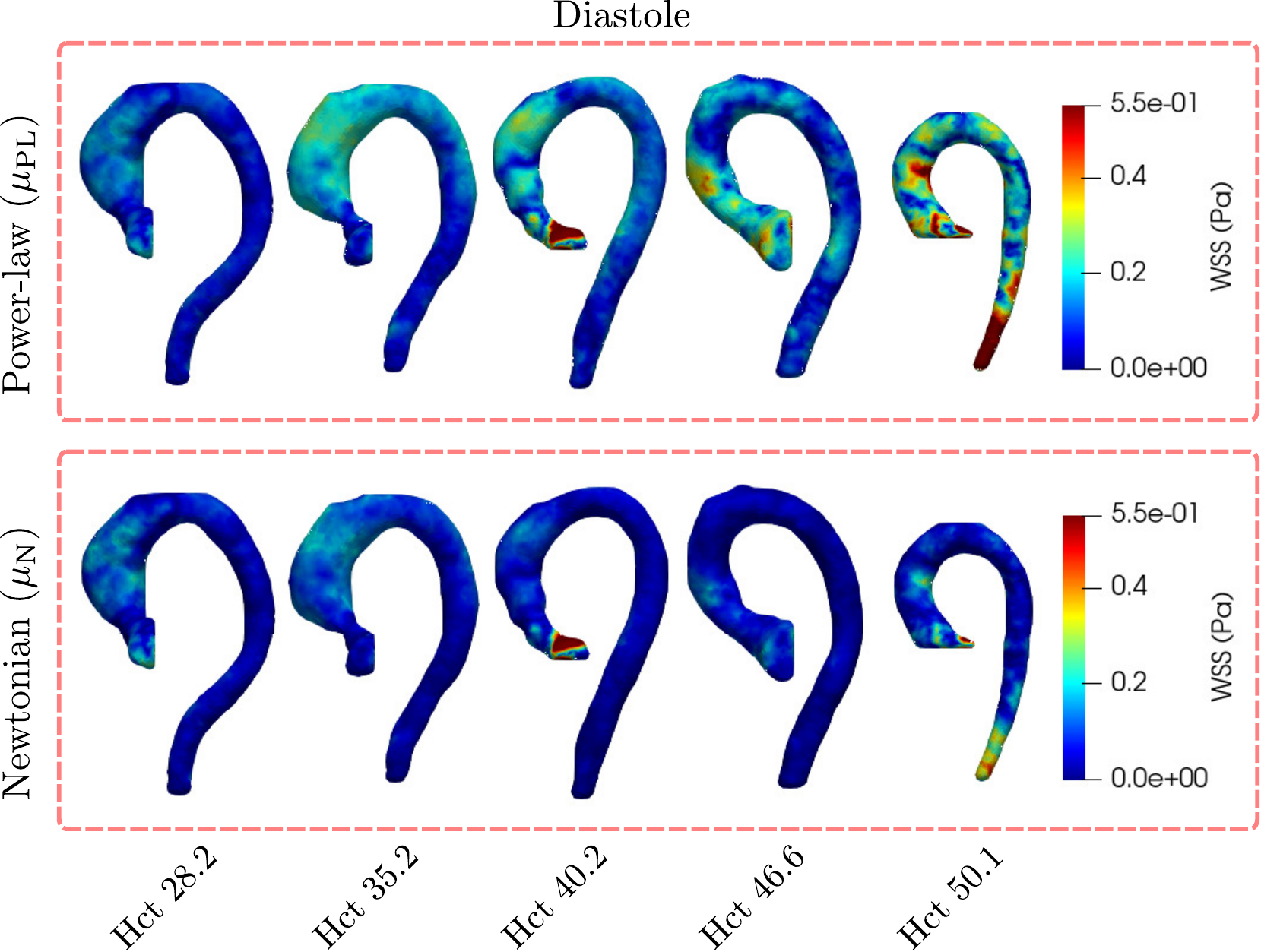}
    \caption{WSS maps at diastole estimated from in-vivo 4D Flow images of patients with HCM at different Hct levels. The estimations shown were obtained using power-law ($\mu_{\mathrm{PL}}$) viscosities and a Newtonian viscosity of $\mu_{\mathrm{N}} = 0.0035$ $\text{Pa}\cdot\text{s}$.}
    \label{fig:wss_maps_invivo_dias}
\end{figure}

Finally, Figures \ref{fig:wss_maps_invivo_sys} and \ref{fig:wss_maps_invivo_dias} show the WSS maps estimated from each patient's data at peak systole and diastole, respectively. The differences described in the previous paragraphs are clearly illustrated in these figures. Additionally, readers are referred to Figure \ref{fig:appendix-3} in \ref{sec:sample:appendix} for the means and standard deviations of WSS, $E_L$, and OSI within each aortic segment.

\section{Discussion}
\label{sec:discussion}
\review{It is noteworthy that for higher Hct values (i.e., higher viscosities), the error in the fitted power-law viscosities is greater (see RMSE values reported in Figure \ref{fig:viscosities-a}). This is worth discussing because, for higher viscosities, there should be less experimental error in the measurements, implying greater certainty about the data. Moreover, this observation could suggest that the power-law model might not be the optimal choice to represent the experimental data. However, as mentioned in Section \ref{ssec:shear-stress-behavior}, this model was chosen based on its practicality: it can be fully defined using 4D Flow MR velocity measurements and a blood test, which is always available during the examination. Therefore, no additional experiments are required to determine the parameters of a more complex rheological model, nor is specialized equipment needed. This makes the power-law model suitable for adoption in any clinical context.}

\review{Also related to the fitting procedure, the adopted methodology for estimating Newtonian viscosities fitted by Hct yields a viscosity value of 0.97 mPa, which is lower than the plasma viscosity reported in the literature, typically ranging from 1 to 1.6 mPa \cite{horner2020, beris2021}. However, as mentioned earlier, the methodology employed in this study was the most appropriate choice given the available blood viscosity measurements across the Hct range considered, while also maintaining the physical consistency of the fit.}

Experiment 1 tested the necessity of a non-Newtonian model when computing key parameters such as those defined in \eqref{eq:hemodynamic-parameters} \review{directly from synthetic velocity measurements obtained from synthetic 4D Flow MR images generated from CFD simulations}. The significance of this experiment lies in assessing whether the Newtonian linear model ($\mu_{\mathrm{NF}}$) could serve as a surrogate for the power-law constitutive model, thereby eliminating the need to account for nonlinear rheology. However, the differences observed in Figure \ref{fig:hem_params_adj_sims} indicate that the two models are neither interchangeable, similar, nor equivalent. In some cases, differences between models reached up to -63\%.  These results suggests that, given the well-established shear-thinning behavior of blood in the literature, a constant viscosity model is insufficient for estimating hemodynamic parameters \review{from 4D Flow velocity measurements}. Furthermore, the findings indicate that flow dynamics cannot be accurately characterized using a constant viscosity, highlighting the need to consider nonlinear models.

Since we have established the necessity of accounting for the nonlinear behavior of blood, Experiment 2 evaluated the extent of the differences in WSS, $\dot{E}_L$, and OSI, \review{estimated from velocity measurements obtained from synthetic 4D Flow MR images,} when using an incorrect (but commonly employed) Newtonian (constant) viscosity ($\mu_{\mathrm{N}}$) compared to power-law viscosities ($\mu_{\mathrm{PL}}$) fitted to Hct values. Results showed significant differences between the estimations made with the two models (see Figures \ref{fig:hem_params_sims}, \ref{fig:wss_maps_sims_sys}, and \ref{fig:wss_maps_sims_dias}), highlighting that using a constant viscosity model can lead to inaccurate predictions of hemodynamic parameters when nonlinear blood rheology is present. This was further supported by Figure \ref{fig:wss_comparison_simulations}, where WSS maps estimated from \review{velocity measurements obtained from synthetic} 4D Flow \review{MR images} demonstrated similar spatial distributions and Hct-level variations compared to their CFD simulation counterparts, although significant underestimations were observed at peak systole.

However, the underestimation found at peak systole aligns with previous studies, which indicate that hemodynamic parameters, particularly WSS, are consistently underestimated when using 4D Flow MRI \cite{peterssonAssessmentAccuracyMRI2012, dyverfeldt4DFlowCardiovascular2015, Sotelo2016, cibisEffectSpatialTemporal2016, rodriguez-palomaresAorticFlowPatterns2018, szajerComparison4DFlow2018, ferdianWSSNetAorticWall2022, cherryImpact4DFlowMRI2022}. Additionally, we observed the same behavior reported in \cite{szajerComparison4DFlow2018} and \cite{cherryImpact4DFlowMRI2022}: WSS is significantly underestimated during systole (around 40\% in our case), with these differences diminishing during diastole. These findings provide confidence that all estimations made from \review{4D Flow MR} images are faithful representations of the ground-truth estimations derived from CFD simulations.

Up to this point, it is important to acknowledge that Experiments 1 and 2 were conducted using only one geometry from a healthy volunteer under various viscosity scenarios. However, we believe that the conclusions discussed in the previous paragraphs remain valid due to the way we formulated our research questions. Firstly, we aimed to determine whether using a linear viscosity model could yield equivalent results to those obtained with a nonlinear rheology. Secondly, if the answer to the first question was negative, we sought to quantify the differences when considering the nonlinear rheology of blood versus the current standard approach. Therefore, to answer both questions, presenting one counterexample is sufficient.

Hemodynamic parameters obtained from in-vivo data in Experiment 3 also exhibited notable differences compared to the results obtained in experiments 1 and 2. However, despite the insightful and promising results, their interpretation must be carefully approached. This is due to several reasons: first, the limited number of data points (five) is insufficient for claiming definitive trends. Second, the range of Hct considered in the in-vivo experiments is smaller than that considered in the simulations (28.2 to 50.1 versus 20 to 70, respectively), and third, the data was obtained from patients with different types of HCM. Therefore, it is a future endeavor of the authors to extend this study to a larger population, including healthy volunteers and patients with various diseases, to obtain trends and statistics that would be clinically meaningful.

The observed differences between estimations using the power-law ($\mu_{\mathrm{PL}}$) and Newtonian ($\mu_{\mathrm{N}}$) viscosities are noteworthy and warrant discussion. As highlighted earlier, hemodynamic parameters are often underestimated in 4D Flow MRI due to resolution-related imaging effects, particularly partial volume effects. This underestimation can be substantial, exceeding 50\% of the ground-truth value during systole for a pixel size of 2 mm \cite{cherryImpact4DFlowMRI2022}. It could be argued that such underestimations might influence the differences observed in this investigation. However, it is important to recognize that these underestimations are inherent in all 4D Flow-derived hemodynamic parameters. Despite this limitation, our study still revealed significant differences and consistent hemodynamic parameters when compared to CFD simulations.

\review{It is worth mentioning that our results obtained from CFD simulations are consistent with findings reported in the existing literature. For instance, we observed that hemodynamic parameters increased with Hct in non-Newtonian simulations, which aligns with the results presented in \cite{khan2024} for carotid artery simulations. However, the same cannot be asserted for hemodynamic parameters estimated from 4D Flow MR images, as, to the authors’ knowledge, this is the first study to thoroughly assess this using both synthetic and in-vivo aortic 4D Flow MR data across a wide range of Hct values. However, the authors acknowledge the valuable contributions of 4D flow studies that incorporate non-Newtonian blood rheologies \cite{cheng4DFlowMRI2019,riva4DFlowEvaluation2021}.}

Regarding the current approach to estimating hemodynamic parameters, it is important to note that the high variability in Newtonian viscosity values reported in the literature could introduce bias into the statistics of viscosity-dependent parameters, potentially limiting their diagnostic utility. Therefore, using a measurable criterion such as Hct to standardize the estimation of hemodynamic parameters would represent a significant improvement. In this context, we believe that introducing a simple shear-thinning viscosity model, such as the power-law, could enhance the accuracy of viscosity-dependent hemodynamic parameter estimation, ultimately leading to a better understanding of cardiovascular diseases.

Having said that, it is important to note that this investigation focused solely on the use of the power-law model. Other viscosity models, such as the Carreau-like \cite{Box2005,Lee2007,Gharahi2016}, Cross \cite{Abugattas2020,Mendieta2020}, and Casson models \cite{Suzuki2021}, were not considered. However, as shown in the literature, the power-law model has produced similar results in terms of flow dynamics and hemodynamic parameters when compared to other models in vessels such as the coronary arteries and the aorta \cite{soulis2008,karimi2014}, \review{although it may not capture phenomena such as the plastic component of blood or inhomogeneous effects}. That said, it remains unclear whether these differences between models could be detected using 4D Flow MRI measurements, given the presence of partial volume effects, flow artifacts, and noise.

A limitation of this investigation is the lack of readily available power-law parameters for a wide range of Hct levels at physiological temperature (e.g., 37$^\circ$C). In most of the literature reviewed for this work, only graphical representations of the adjusted curves were provided, making the task of extracting the values of $m$ and $n$ complicated. This challenge led to the development of the methodology used in this study to calculate these values for any Hct level within the range of 20 to 70 (as described in Section \ref{ssec:newtonian visco estimation}) based on discrete Hct data from a single article \cite{wells_influence_1962}. This limitation underscores the need for standardizing blood rheology research by providing both the parameters of the adjusted models and the corresponding measurements.

All previous experiments and discussions assumed that the Hct level is known for every patient at the time of the 4D Flow exam. While this assumption may not always hold, it is reasonable to expect that, in most cases, this data will be available, as CBCs are routinely performed in patients with heart failure and related comorbidities \cite{heidenreich2022AHAACC2022}. Additionally, CMR exams often use gadolinium contrast agents, which are contraindicated in patients with severe renal impairment, necessitating a CBC to assess renal function. Furthermore, this assumption is even more plausible considering that CMR exams are typically conducted later in the diagnostic process, primarily due to their higher cost. However, this raises a related question: is the CBC still valid at the time of the 4D Flow exam?

Many efforts have been made to address this question in various contexts, including MRI. For instance, studies have shown that not only Hct but also several other hematological parameters can vary with posture \cite{jacobPosturalPseudoanemiaPostureDependent2005}, throughout the day \cite{sennelsDiurnalVariationHematology2011, hilderinkWithindayBiologicalVariation2017}, and seasonally \cite{statlandEvaluationBiologicSources1978, maesComponentsBiologicalIncluding1995, thirupHaematocrit2003}. Other sources of variability include pre-analytical factors such as fasting condition, venous stasis, and exercise, which are usually standardized to reduce variation \cite{thirupHaematocrit2003}.

Considering these factors, the reported variations in Hct have been demonstrated not to affect the estimation of CMR-derived parameters directly dependent on Hct, such as extracellular volume \cite{suTimelyAssessmentHematocrit2020}. Thus, we hypothesize that these variations would not induce significant differences in hemodynamic parameters estimated from 4D Flow data. However, this remains an open question.

\section{Conclusion}
\label{sec:conclusion}
Our study highlights the importance of accounting for the nonlinear behavior of blood when estimating hemodynamic parameters from 4D Flow MRI using available CBC data. The conventional use of Newtonian or Newtonian-fitted viscosities is insufficient for capturing the complex dynamics of blood flow, particularly in major vessels such as the aorta. This research demonstrates that using the simple power-law model as the ground truth leads to significant differences in both simulations and synthetic 4D Flow MRI data when compared to linear Newtonian (or Newtonian-fitted) models.

\review{Since blood is intrinsically non-Newtonian, accounting for this behavior during the processing of 4D Flow MR images may enhance the accuracy and realism of hemodynamic parameter estimation. In this context, modeling viscosity using a power-law formulation emerges as a reasonable and straightforward alternative. Adopting this simple yet impactful methodological change could significantly improve the diagnosis and assessment of cardiovascular diseases through 4D Flow MR examinations, as health conditions involving variations in Hct would be directly reflected in the estimated hemodynamic parameters.}
 
During the development of this study, a key challenge encountered was the lack of readily available power-law parameters for a wide range of Hct levels under physiological conditions. \review{Although we highlight the existence of numerous notable studies assessing the role of Hct in blood rheology (see, for instance, \cite{walburn_constitutive_1976,quemada1981,baskurt2003,horner2020,beris2021}), they were not applicable to this research due to limitations such as a narrow Hct range, the use of different viscosity models, non-physiological measurement conditions, or the unavailability of raw data}. This limitation highlights the need for future research aimed at developing comprehensive databases of power-law parameters, as well as those for other rheological models, across varying Hct values. Such resources would enhance the accuracy of hemodynamic parameter estimation and promote the standardization of blood rheology research. In this regard, the methodology described in Section \ref{ssec:newtonian visco estimation} is freely available at \href{https://www.github.com/hmella/power-law-parameters}{www.github.com/hmella/power-law-parameters}.

\review{We also identified other limitations of this study. For instance, viscosity models such as the Carreau-like, Cross, and Casson were not considered in this investigation. Additionally, it remains unclear whether CBC exams conducted on a different day from the 4D Flow MR exam would impact the estimation of hemodynamic parameters, although some studies in other MRI contexts suggest that this may not be the case. Finally, the number of in-vivo datasets analyzed in this study is limited.} These limitations motivate our future work, which will focus on three main areas. First, we will investigate using other purely viscous rheological models, such as Carreau-like, Cross, and Casson, for estimating hemodynamic parameters \review{from velocity measurements obtained using 4D Flow MR imaging}. Second, we will assess the impact of timely Hct measurements on estimating hemodynamic parameters \review{from 4D Flow MR images}. Finally, we will conduct a statistical analysis across a wide range of Hct values to evaluate the robustness and clinical applicability of the proposed \review{methodology} in both volunteers and patients.

\section{Acknowledgments}
The authors sincerely acknowledge the financial support provided by ANID through the following projects: Fondecyt de Iniciación en Investigación \#11241098 (HM) and \#11200481 (JS), Fondecyt Regular \#1210156 (EC) and \#1250287 (EC and FG), and the Millennium Science Initiative Program ICN2021\_004 (JS). HM and JS also acknowledge the Department of Medical Imaging and Radiation Sciences at Monash University. FG also acknowledges the funding provided by the project DI VINCI PUCV 039.731/2025. The authors would like to express their sincere thanks to the Weierstraß-Institut für Angewandte Analysis und Stochastik (WIAS) for providing access to their servers for running CFD simulations.



\newpage
\appendix
\setcounter{figure}{0}
\section{Hemodynamic parameters for each aortic segments}
\label{sec:sample:appendix}

\begin{figure*}[!hbt]
    \centering
    \includegraphics[width=\textwidth]{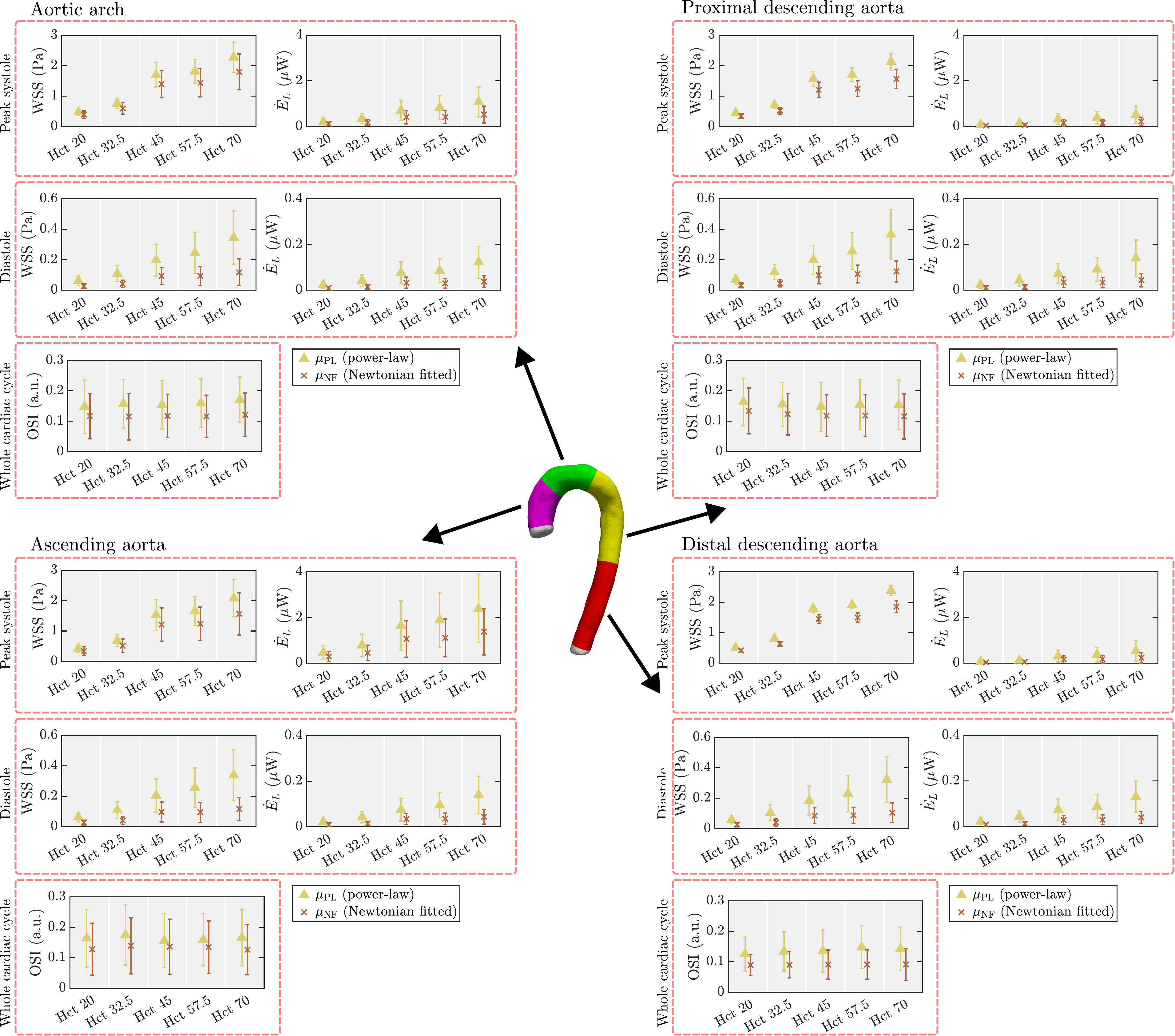}
    \caption{Mean values and standard deviations of WSS, $\dot{E}_L$, and OSI within each aortic segment, estimated from simulated 4D Flow images generated from CFD simulations using power-law ($\mu_{\mathrm{PL}}$) and Newtonian fitted ($\mu_{\mathrm{NF}}$) viscosities. The estimations were performed using viscosity models and values consistent with those used in the CFD simulations that generated the synthetic 4D Flow images. Results are presented for both peak systole and diastole. Higher values and greater variability are observed in the ascending aorta due to the flow dynamics. The same differences shown in Figure \ref{fig:hem_params_adj_sims} are clearly depicted here for each segment.}
    \label{fig:appendix-model-interrogation}
\end{figure*}

\begin{figure*}[!hbt]
    \centering
    \includegraphics[width=\textwidth]{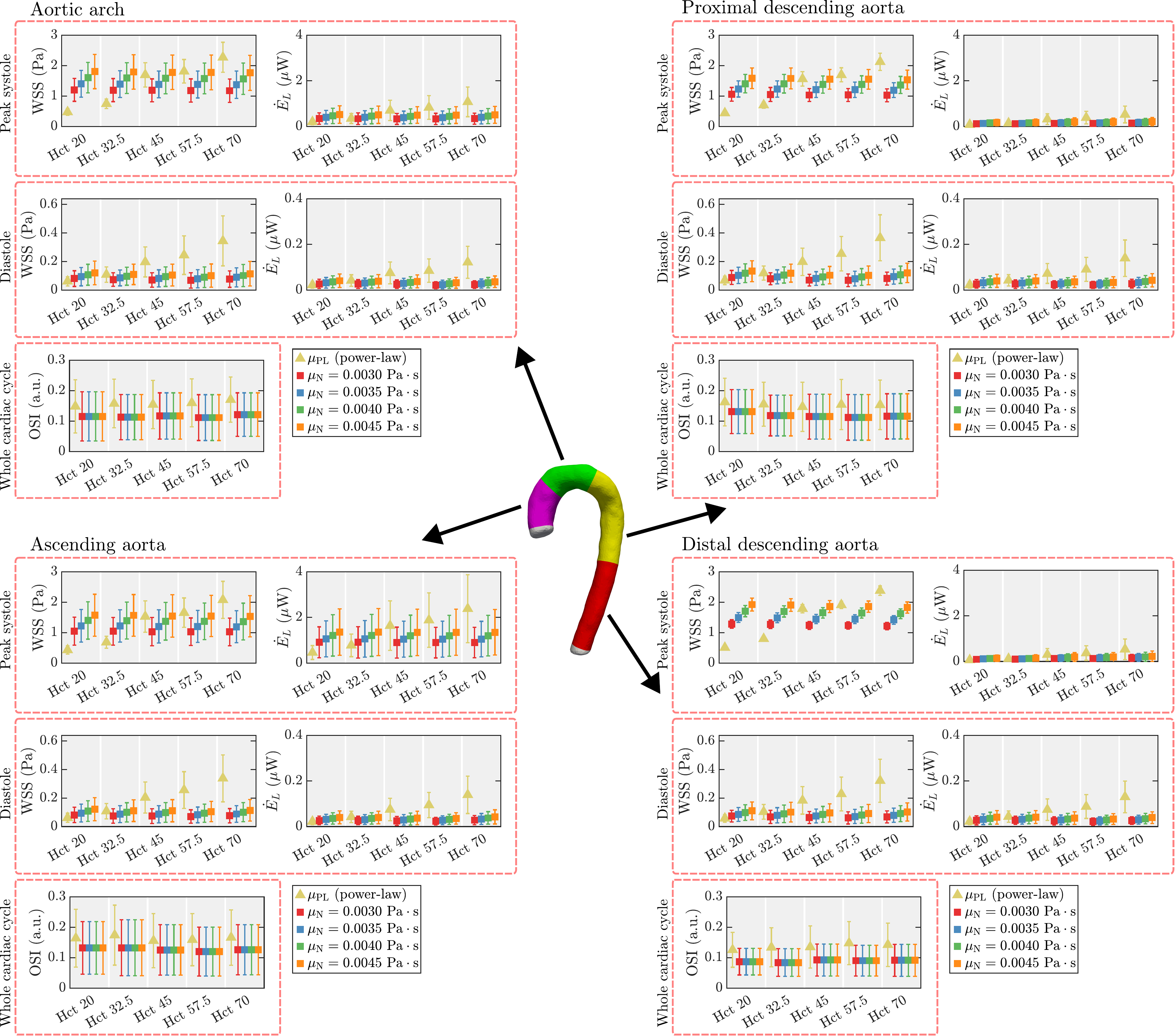}
    \caption{Mean values and standard deviations of WSS, OSI, and $\dot{E}_L$ within each aortic segment estimated from synthetic 4D Flow images generated by CFD simulations obtained using power-law viscosities. Hemodynamic parameters were estimated using both power-law ($\mu_{\mathrm{PL}}$) and state-of-the-art Newtonian ($\mu_{\mathrm{N}}$) viscosities. Results are presented for both peak systole and diastole. Higher values and greater variability are observed in the ascending aorta due to the flow dynamics. The same differences shown in Figure \ref{fig:hem_params_sims} are clearly depicted here for each segment.}
    \label{fig:appendix-2}
\end{figure*}

\begin{figure*}[!hbt]
    \centering
    \includegraphics[width=\textwidth]{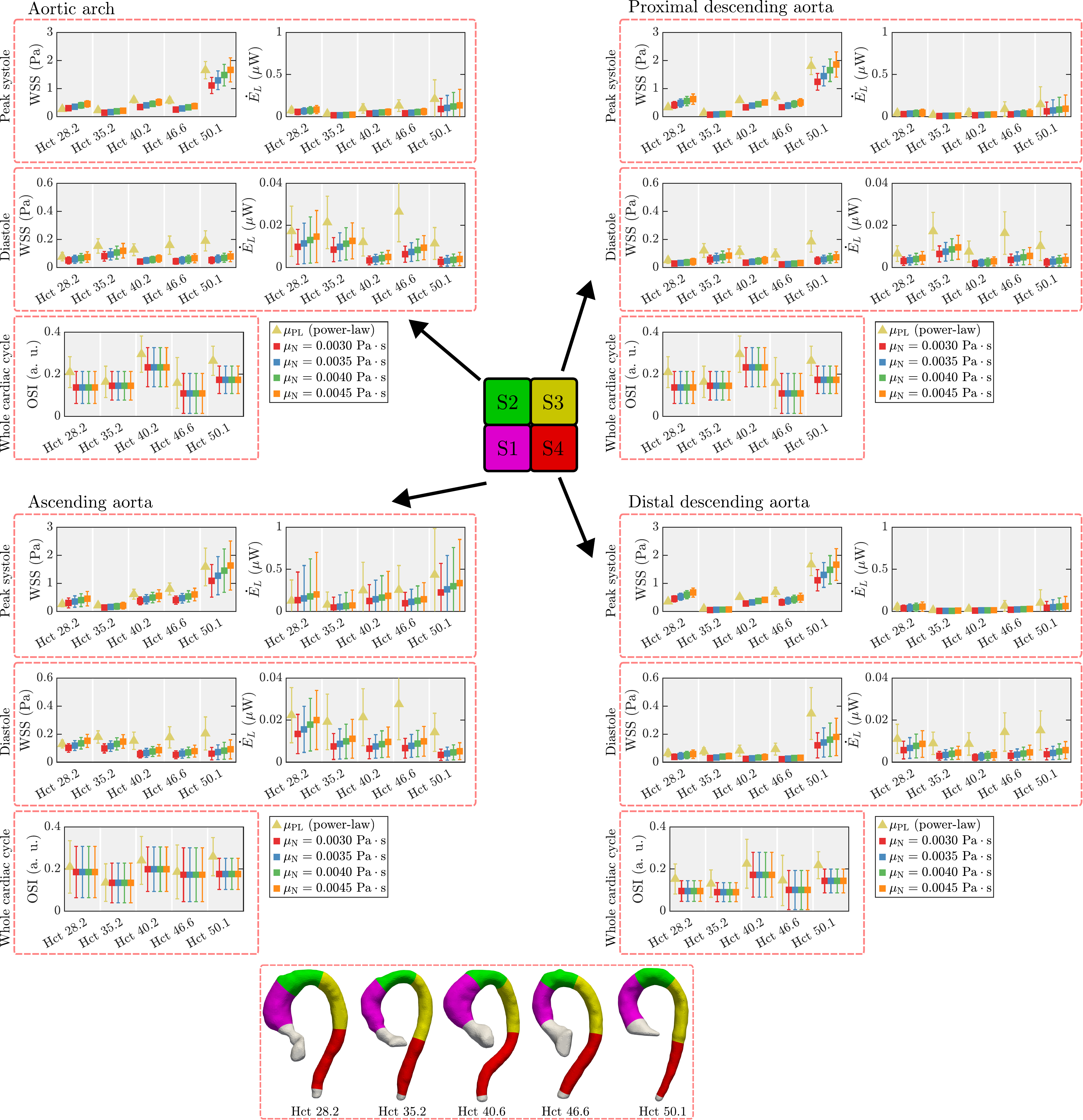}
    \caption{Mean values and standard deviations of WSS, OSI, and $\dot{E}_L$ within each aortic segment estimated from in-vivo 4D Flow images of HCM patients. Hemodynamic parameters were estimated using both power-law ($\mu_{\mathrm{PL}}$) and state-of-the-art Newtonian ($\mu_{\mathrm{N}}$) viscosities. Results are presented for both peak systole and diastole. Higher values and greater variability are observed in the ascending aorta due to the flow dynamics. The same differences shown in Figure \ref{fig:hem_params_invivo} are clearly depicted here for each segment.}
    \label{fig:appendix-3}
\end{figure*}

\end{document}